# Modelling temporal dynamics of suicidal ideation and behaviour across pre- to early adolescence using a Markov framework

Sieun Lee[1,2], Ben Cardoen[3], Marianne Etherson[4], Nitish Jawahar[1,5], Ellen Townsend[6], Kapil Sayal[7,8], Peter Fonagy[9], Aja Murray[10], Joanna Lockwood[1,11], Ayan Mahamud[12], Chris Hollis[1,8,11], Rory O'Connor[4], Dorothee Auer[1,5,13] on behalf of the Digital Youth Research Programme

[1]Mental Health & Clinical Neurosciences, School of Medicine, University of Nottingham, Nottingham, United Kingdom

[2]School of Social Sciences, University of Manchester, Manchester, United Kingdom

[3]School of Mathematics, University of Birmingham, Birmingham, United Kingdom

[4]Suicidal Behaviour Research Lab, School of Health and Wellbeing, University of Glasgow, Glasgow, United Kingdom

[5]Sir Peter Mansfield Imaging Centre, Nottingham, United Kingdom

[6]School of Psychology, University of Nottingham, Nottingham, United Kingdom

[7]School of Medicine, University of Nottingham, Nottingham, United Kingdom

[8]Institute of Mental Health, Nottinghamshire Healthcare NHS Foundation Trust, Nottingham, United Kingdom

[9]Research Department of Clinical, Educational, and Health Psychology, University College London, London, United Kingdom

[10]Department of Psychology, University of Edinburgh, Edinburgh, United Kingdom

[11]NIHR MindTech HealthTech Research Centre, School of Medicine, University of Nottingham, Nottingham, United Kingdom

[12]Sprouting Minds (Young People Advisory Group – Digital Youth Programme), University of Nottingham, Nottingham, United Kingdom

[13]NIHR Nottingham BRC, University of Nottingham, United Kingdom

*Corresponding author: Sieun Lee (sieun.lee1@nottingham.ac.uk)

## Abstract

Understanding the dynamics of suicidal ideation and behaviour in youth and the factors associated with transitions from thoughts to behaviours is critical for early identification, monitoring, and prevention. Using longitudinal self-report data from the Adolescent Brain Cognitive Development (ABCD) Study (n = 11,864) spanning ages 9 to 13 years, we developed a time-inhomogeneous discrete-time Markov chain framework to model transitions across eight states defined by suicidal ideation and behaviour and the co-reported presence or absence of non-suicidal self-injury (NSSI). The framework enables inference of year-to-year and multi-year transition probabilities, quantification of transition stability and uncertainty, and statistical comparison of transition likelihoods. We identified structured but developmentally changing transition patterns, including a generally high probability of recovery to a no-report state following reports of suicidal ideation or behaviour. Co-reported lifetime NSSI marked a distinct trajectory profile characterized by both greater risk and lower predictability: children with NSSI were more likely to transition to or persist in suicidal behaviour, were less likely to recover to a no-report state, and exhibited greater uncertainty in their transition likelihoods compared to those reporting suicidal ideation and/or behaviour alone. These findings suggest that NSSI marks not only elevated risk but also greater trajectory instability during pre- to early adolescence. By estimating probabilities of escalation, remission, persistence, and fluctuation in longitudinal cohort data, the framework provides a systematic, interpretable, and scalable approach to characterising suicidal trajectory dynamics in large, sparse mental health datasets, with potential to inform future research on monitoring, early prevention, and time-varying risk in youth mental health.

## Main

Suicide is a leading cause of death among adolescents, with suicidal ideation and self-injurious behaviour often first emerging in late childhood or early adolescence.[1–3] A recent systematic review reported that the lifetime prevalences of suicidal ideation, suicidal attempts, and non-suicidal self-injury (NSSI) among preadolescent children (age <13 years) were 15.1%, 2.6%, and 6.2%, respectively.[4] According to the ideation-to-action framework of suicide, the development of suicidal ideation and the progression to suicide attempts are distinct processes, each influenced by different mechanisms and predictors.[5] Understanding the developmental pathways and risk factors associated with this transition is critical for improving early detection and prevention efforts.

However, the mechanisms underlying these transitions remain poorly understood because of inconsistent findings, limited longitudinal data, and heterogeneity in study populations.[6–8] NSSI, defined as self-inflicted harm without suicidal intent,[9] is a robust predictor of later suicide attempt.[10] Yet the processes by which NSSI contributes to suicidal behaviour remain an active area of investigation.[10,11] Most existing research has focused on cross-sectional associations or modelling longitudinal patterns as linear or continuous, which may overlook the complex and unstable nature of suicidal ideation and behaviour, especially in youth.

Recent longitudinal studies, including those leveraging the large-scale Adolescent Brain Cognitive Development (ABCD) study,[12] have shown that suicidal ideation and behaviour in youth are often variable and nonlinear, with substantial within-individual variation over time.[13] Though important insights have been gained from longitudinal modelling of suicidal ideation and behaviour,[14,15] the vast majority of research to date has relied on traditional modelling approaches that depend on assumptions of continuous change, metric properties, or pre-specified trajectories, creating corresponding limitations to our understanding of the dynamics of suicidal ideation and behaviour and NSSI in this critical developmental period.[13,16] Assessments in large community cohorts, such as ABCD Study, are further constrained by sparse measurement timing, reporting inconsistencies, and categorical symptom frameworks.

Recent work has approached self-harm as sequential rather than static by modelling the ordered patterning of thoughts, feelings, events, and behaviours leading up to self-harm episodes. Using the Card Sort Task for Self-harm (CaTS), lag sequential analyses have identified transitions and highlighted antecedent factors including hopelessness and wish to die.[17] Extension of this work to high-risk groups[18] and online adaptations in combination with Indicator Wave Analysis[19] has further demonstrated structured transitions and clinically salient proximal markers, and how the temporal prominence of these indicators shifts across time leading up to a self-harm act.

From a developmental psychopathology perspective, the instability in NSSI and suicidal ideation and behaviour observed in preadolescence is not merely a stochastic process but reflects a dynamic system undergoing reorganization.[20,21] During this period, rapid changes in affect regulation, interpersonal stress, and self-injury functions create feedback loops that can transiently amplify or dampen self-injurious and suicidal thoughts and behaviours.[22,23] Such instability may represent a developmental signature of emerging self-organization, as affective and behavioural systems calibrate to internal and external stressors. Capturing these dynamics and their variability over time is critical for understanding the developmental architecture of risk and resilience.

To address these challenges, we propose a probabilistic framework based on discrete-time Markov chains (DTMC)[24] for modelling longitudinal transitions between discrete states of suicidal ideation and behaviour and NSSI in pre- to early adolescence. Whereas prior sequential work, such as CaTS, characterises the ordered micro-temporal build-up to self-harm episodes, the proposed approach aims to estimate year-to-year and multi-year transition structure in a community cohort, where observations are typically large in number but categorical, temporally sparse, and noisy. Markov models allow for flexible, data-driven estimation of transition probabilities between defined states without requiring assumptions about linearity or pre-specified latent trajectory shape. Crucially, we extend standard DTMC use by i) allowing time-inhomogeneous (age-dependent) transition matrices, ii) quantifying uncertainty in rare transitions using bootstrap confidence intervals and entropy-based measures of predictability, and iii) enabling queries of custom multi-year trajectory probabilities so that clinically meaningful patterns such as escalation, persistence, remission, and fluctuation can be evaluated directly. The model therefore provides an empirical, developmentally grounded way to operationalise ideation-to-action transitions as estimable probabilities over specified windows, while preserving the observed volatility of preadolescent presentations.

Here, we apply this framework to the ABCD Study data spanning a follow-up of four years (age 9–14 years) available at the time of this study, focusing on transitions between eight mutually exclusive suicidal and NSSI states defined by the presence or absence of NSSI, passive or active suicidal ideation, and suicidal behaviour. We first describe patterns in prevalence, co-occurrence, and child-parent reporting discrepancies over time. These results highlight the increasing prevalence and intensity of suicidal ideation and behaviour during this pre- to early-adolescent developmental window, a period that marks the onset of a pronounced rise in suicidal thoughts and behaviours across adolescence and immediately precedes the mid-adolescent peak in self-harm.[21,25] We then use the Markov framework to quantify year-to-year and multi-year transition likelihoods, assess trajectory uncertainty, and evaluate the specific risk associated with co-occurrence of NSSI.

# Results

## Descriptive statistics on preadolescent suicidal ideation and behaviour and NSSI

### Prevalence of suicidal ideation and behaviour and NSSI

Table 1a and Figure 1a show the prevalence of lifetime reports of NSSI and suicidal ideation and behaviour from baseline (age 9-10) to 4-year follow-up (age 13-14). The prevalence of all five ideation-or-action states (NSSI, passive ideation, active ideation, preparatory action, and suicidal attempt) generally increased during this period, especially from age 11-12. NSSI and ideation were more commonly reported than suicidal behaviours, but suicidal behaviours increased more sharply compared to the baseline.

Because children often reported more than one type of suicidal ideation or behaviour, prevalence was also analysed by individual according to their most advanced reported state (Figure 1b (top)). At age 9-10, approximately 9% of the children reported lifetime suicidal ideation or behaviour. This rate decreased slightly to ~8% at age 11-12 and then rose to over 12% by age 13-14. Active suicidal ideation and suicidal behaviour increased more than passive ideation, such that by age 13-14 a larger proportion of children reported more advanced stages of suicidal ideation and/or behaviour. This was in part due to more children at older ages reporting active ideation or behaviour as their first-time report (Fig. 1c (bottom)).

### Co-occurrence of NSSI, suicidal ideation, and suicidal behaviour

Twenty-five percent of the children who reported any lifetime NSSI or suicidal ideation or behaviour at age 9-10 reported more than one type of self-injurious thoughts or behaviour.

This co-occurrence increased as the children got older to 44% at age 13-14. Table 2 and Extended Data Figure 1 provide the detailed breakdown.

*Co-occurrence of NSSI and suicidality*

Among children who reported NSSI, co-reporting suicidal ideation or behaviour increased from 41% at age 9-10 to 65% at age 12-13 (Table 2a). Likewise, among those who reported suicidal ideation or behaviour, co-reporting NSSI increased from 28% to 47%. The percentage of co-reported NSSI also increased from passive ideation to active ideation to suicidal behaviour. Among those reporting passive suicidal ideation, 18-21% co-reported NSSI, whereas 30-42% of those reporting active suicidal ideation and 52-74% of those reporting suicidal behaviour co-reported NSSI.

*Co-occurrence of suicidal ideation and suicidal behaviour*

At age 9-10, 48% of children who reported lifetime active suicidal ideation also co-reported lifetime passive suicidal ideation, increasing to 66% by age 13-14. (Table 2b). Only a small minority of children reporting suicidal attempt (8-14%) co-reported preparatory actions. Most suicidal attempt incidences co-reported ongoing or preceding suicidal ideation, from 85% at age 9-10 to 96% at age 13-14.

**Child-parent reporting discrepancy**

In Table 1b and Figure 1c, the parent reports showed similar prevalence and patterns to child reports, with slightly lower values and a sharper increase in suicidal attempt by ages 13-14. Table 3 summarizes the child-parent discrepancies. Most of the reports across outcomes and age were not corroborated between the child and parent: 70-99% of reports from either child or parent were reported only by the child or the parent. This discrepancy was generally higher for suicidal behaviours (Prep action (91-99%) and attempt (79-90%)) than NSSI (70-91%) or suicidal ideation (72-88%). There was some decrease in the discrepancies over time. At age 9-10, 76-100% of reports were endorsed by the child or parent alone, and by age 13-14 this proportion fell to 47-85%. However, even at this age, only a minority of NSSI or suicidal ideation or behaviour reports (10-30%) were reported by both the child and parent. The remaining analysis in this paper is based only on the child reports, since they directly reflect the subjective internal experiences of the youth themselves that may be unknown to caregivers.

## Longitudinal transitions and trajectories in NSSI and suicidal ideation and behaviour

### Year-to-year transition probabilities

The year-to-year transition probability matrices in Figure 2 delineate the full set of 1-year transition probabilities between NSSI, suicidal ideation, and suicidal behaviour states across four time points between ages 9-13 years. States are 1 = no non-suicidal self-injury (NSSI) or suicidal ideation or behaviour; 2 = NSSI only; 3 = passive suicidal ideation without NSSI; 4 = passive suicidal ideation with NSSI; 5 = active suicidal ideation without NSSI; 6 = active suicidal ideation with NSSI; 7 = suicidal behaviour without NSSI; 8 = suicidal behaviour with NSSI. Rows are origin states at year $t$ and columns are destination states at year $t + 1$. The colour encodes direction of transition (green = to a lower state; red = to a higher state; orange = persisting in the same state), and opacity encodes probability (more opaque = more likely). Cell annotations report the mean transition probabilities with standard deviation. This representation enables visual comparison of transition patterns across states and years. For example, in the lowest matrix (Year 3 to 4), children reporting active suicidal ideation with NSSI at age 11-12 (Row 6) had a 6% probability of transitioning to suicidal behaviour in the following year (Column 7 and 8), compared with 4% among those reporting active ideation only (Row 5, Column 7 and 8).

Transition probability patterns were relatively consistent over time, with developmental shifts in risk and recovery as children got older. A prominent feature across all years was the relatively high likelihood of remission or recovery (absence of report of NSSI, suicidal ideation, or suicidal behaviour; Column 1) which was the most likely next-year outcome for all originating states except suicidal behaviour combined with NSSI (Row 8). This recovery probability declined with increasingly advanced originating state. The transition matrices also reveal structured heterogeneity linked to the presence of NSSI (indices 4, 6, and 8). A checkered pattern in row- and column-wise alternation indicates that co-occurrence of NSSI modifies transition probabilities in consistent ways. Specifically, histories of NSSI were associated with significantly lower probabilities of remission and significantly higher probabilities of escalating to or persisting in suicidal behaviour.

For an alternative interpretation of the data, we additionally constructed simpler transition matrices based on five composite states: 1 = no NSSI or suicidal ideation or behaviour; 2 = NSSI only; 3 = suicidal ideation without NSSI; 4 = suicidal ideation with NSSI; and 5 = suicidal behaviour. (Figure 3). The matrices show overall reduction in the recovery transition probabilities (first column) and increase in the suicidal behaviour transition probabilities (last columns) with increasing age and presence of NSSI.

**Year-to-year transition uncertainties**

To evaluate the predictability of year-to-year transitions, we computed the Shannon entropy of all year-to-year transitions, where entropy of 0 indicates no uncertainty in prediction and a higher entropy indicates a greater uncertainty. As shown in Figure 4, uncertainty increased from passive suicidal ideation to behaviour, with the highest entropy observed in suicidal behaviour (States 7 and 8). Across all states and years, those with co-reported NSSI (States 4, 6, and 8) exhibited significantly greater transition uncertainty than their counterparts without NSSI (States 3, 5, and 7), indicating a more unstable and less predictable profile. Entropy also increased over time, which reflects growing heterogeneity in trajectories as children aged.

**Statistical comparisons and distilling complex patterns**

In Figure 2, we noted a checkered pattern in the transition probability matrices, indicating that co-reported lifetime NSSI consistently influenced suicidal transitions. Figure 5 demonstrates how transition probabilities and uncertainties can be used to investigate such specific patterns. Figure 5a plots 1-year probabilities of escalating to or persisting in suicidal behaviour from passive ideation (left) and active ideation (middle) and suicidal behaviour (right) with 95% confidence interval bands. In all comparisons, children with NSSI showed significantly higher (non-overlapping) probabilities of transitioning to or persisting in a suicidal behaviour state. Figure 5b plots 1-year transition probabilities to recovery (no report) from passive ideation (left), active ideation (middle) and suicidal behaviour (right). Here, in all comparisons, children with NSSI were significantly less likely to recover. The generally wider confidence intervals for NSSI states also align with the higher entropy shown in Figure 4.

Figure 6 demonstrates an example of a custom multi-year trajectory query: "*What is the likelihood that a child reporting passive suicidal ideation at age 9-10 will transition to suicidal behaviour within the next three years, and how does this differ for children co-reporting NSSI?*" The right panel shows that children with passive ideation at age 9-10 years and no history of NSSI have an ~7.8% probability of transitioning to suicidal behaviour within three years. In contrast, those co-reporting NSSI face a significantly higher probability of ~11%. The left panel compares the associated uncertainty, quantified via sequence-level entropy. Trajectories originating from passive ideation combined with NSSI exhibited consistently higher uncertainty than those without NSSI.

## Discussion

Using a time-inhomogeneous discrete-time Markov framework, we quantified how year-to-year NSSI, suicidal ideation, and suicidal behaviour trajectories evolved from age 9 to 12 in the ABCD study dataset. At the cohort level, NSSI, suicidal ideation, and suicidal behaviour reports increased over time, with a sharper rise in active suicidal ideation and suicidal behaviour from age 11-12 onwards. Co-reporting of multiple types of ideation or behaviour became more common with age. Child- and parent-reports broadly showed similar patterns but low concordance, particularly for suicidal behaviours.

Extending beyond static cohort-summary statistics, the Markov chain framework provided a structured account of year-to-year transition dynamics and uncertainty of individual trajectories. We found that trajectories became more heterogeneous and less predictable with age, with uncertainty rising most from age 11-12 and varying systematically by the origin state. Across years, remission was the most common next-year outcome, but its likelihood declined with age and with more advanced states of suicidal ideation or behaviour, while probabilities of persisting in or escalating to suicidal behaviour increased. Superimposed on these dynamics, a history of NSSI consistently shifted trajectories toward greater risk and volatility, with higher likelihood of escalation to or persistence in suicidal behaviour, lower likelihood of remission, and higher uncertainty.

Common limitations in community data such as ABCD Study are relatively coarse temporal resolution and limited qualitative characterisation. Even with these constraints, our multi-state Markov framework revealed patterns in pre- to early adolescent suicidal ideation and behaviour, showing how their transitions were dependent on stage, age, and NSSI, consistent with established theories and evidence on suicide, self-harm, and developmental gradients.[5,6,10,21,25] Across all ages, the probability of escalation to or persistence in suicidal behaviour increased stepwise from passive ideation to active ideation to behaviour, while remission likelihood declined in parallel. The structure suggests systematic dynamics beyond noise in this young cohort. The Markov approach thus operationalises the conceptual progression in ideation-to-action frameworks[24,25] as state- and age-specific transition probabilities, providing interpretable empirical risk estimates. The framework accommodates heterogeneous, fluctuating dynamics of remission, escalation, and persistence across this developmental window, rather than reducing them to a static outcome or a monotonic pattern of worsening or improvement.

The likelihood of remission in the following year ranged from 73% for passive ideation without NSSI at age 9-10 to 26% for suicidal action combined with NSSI at age 11-12, suggesting both optimism and urgency. Although any level of self-injurious thoughts and behaviours is concerning, there was evidence that remission is common even among those who had previously endorsed suicidal thoughts or behaviours. Nonetheless, an urgent response is required because remission declines with increasing age and with more advanced states, highlighting early adolescence as a period of rapidly changing transition dynamics.

One possible interpretation is that suicidal ideation reported at younger ages may be more episodic and context-dependent[13,21], and annual measurement may compress within-year onset and remission into a single observation. In addition, because the ABCD Study includes safety protocols when risk is endorsed[26], the high remission rate could reflect, in part, a combination of natural fluctuation and greater access to monitoring and support following disclosure.

The increased likelihood of children co-reporting NSSI transitioning to or persisting in suicidal behaviour, and their decreased likelihood of remission, may be explained by several mechanisms. NSSI is a well-established predictor of suicide attempt risk.[27–29] Co-occurrence of NSSI with suicidal ideation or behaviour can indicate more chronic and intense stress or distress (e.g. emotion dysregulation, trauma, interpersonal stressors).[30,31] This is indirectly indicated in our results: even at age 9-10, more than half of the children reporting suicidal behaviour co-reported NSSI, compared to 18.4% and 29.7% among the children reporting suicidal ideation. From the ideation-to-action perspective, NSSI can increase capability to enact suicidal behaviour.[32,33] Moving from ideation to attempts requires overcoming fear and pain barriers, and repeated self-injury is associated with habituation to pain and bodily harm, thereby lowering inhibition.[32,34] NSSI may also sustain or increase escalation risk when it temporarily regulates intolerable affect and then fails to provide relief with increased and sustained distress.[35–37]

An important nuance in our findings is that the NSSI-only group without suicidal ideation or behaviour showed similar escalation and remission probabilities as the passive suicidal ideation-only group, while active suicidal ideation fared worse than both. This suggests that NSSI episodes, as captured by annual ABCD K-SADS categories, represent a heterogeneous group, where a substantial proportion may be reporting actions that are transient without sustained harm or suicidal intent. In ideation-to-action framing, suicidal behaviour requires intent with a capability or volitional factor, and in this developmental window active ideation - i.e. a stronger intent - may be a stronger marker for escalation than NSSI by itself. The results, however, point to the consistent role of NSSI as an amplifier of risk and volatility when it co-occurs with suicidal ideation or behaviour.

Beyond transition probabilities, we quantified the uncertainty of transitions using Shannon entropy. Entropy increased with age, state progression, and NSSI, such that children in more advanced states, at older ages, and with co-reported NSSI were less predictable in their transitions. This gradient of increasing uncertainty is in parallel with increased likelihood of escalating to or persisting in suicidal action and decreased likelihood of remission. The results reflect a coherent pattern in how age, state, and NSSI shape the direction and predictability of suicidal trajectories.

More broadly, explicit quantification of uncertainty alongside transition probabilities addresses growing recognition that conventional suicide risk factors have limited predictive utility, and their predictive value varies across individuals and over time.[13,16,38] Retaining and characterising this volatility, rather than smoothing it away, may be essential for understanding youth suicidal ideation and behaviour in the context of neurodevelopment. Early adolescence is characterised by ongoing maturation and reconfiguration of the brain's default-mode and mentalizing systems that support self-referential thought and autobiographical evaluation – processes closely tied to rumination and self-appraisal.[39–41] The increasing entropy with age may reflect a growing heterogeneity in social and developmental context, alongside risk processes that become increasingly context-dependent and time-varying.[42] Adolescents show heightened sensitivity to social evaluation and peer feedback, supported by strong responsivity in socio-affective valuation circuitry.[43,44] This social reorientation can amplify context dependence and volatility in affect and behaviour as peer contexts rapidly change during the transition into adolescence.

In practice, these findings point toward a more developmentally calibrated and uncertainty-informed approach to monitoring suicidal ideation and self-harm in children. Rather than considering apparent recovery as grounds for discharge or endorsement as benign, it may be beneficial to track both the direction and volatility of a child's trajectory over time. A child transitioning from active ideation with NSSI to a no-report state, ostensibly a positive outcome, but with high trajectory entropy, warrants continued monitoring. Such uncertainty-informed triage would complement conventional risk level assessment in guiding screening frequency, follow-up intensity, and targeted preventive intervention, especially in school or primary care settings where clinical decisions must often be made with limited information.

These considerations are especially pertinent for children at elevated baseline risk, for example, those with a history of adverse childhood experiences or early pubertal onset, for whom the window for early intervention may be especially narrow.

Initially, we explored several analytic approaches to characterize longitudinal trajectories of suicidal ideation and behaviour and NSSI and observed substantial fluctuations and heterogeneity in the data. Yet the temporal trends did not appear random, calling for methods that can reveal data-driven patterns without discarding the inherent variability as noise.

Longitudinal mental health data in community samples are often categorical, temporally sparse, and volatile, posing challenges to conventional trajectory modelling that assumes continuous change, metric properties, or pre-specified latent trajectories. The bootstrapped time-inhomogeneous discrete-time Markov chain (DTMC) framework presented here offers several advantages: first, the method is computationally simple, interpretable, and reproducible, with the transition matrices providing a clear and intuitive overview of cohort-level dynamics. Second, it has minimal assumptions and requirements – no predefined trajectories, latent states, or filtering —, which is particularly useful in a community-ascertained sample where data acquisition is infrequent and clinically significant outcomes, such as suicidal behaviours, are rare. Finally, the model naturally incorporates uncertainty, allowing variability in rare outcomes to be represented as part of the probabilistic structure rather than being filtered out.

Building on the DTMC framework, we additionally implemented a query system that allows researchers and clinicians to evaluate the likelihood and uncertainty of custom-defined longitudinal trajectories using a flexible pattern-matching syntax (regular expression) that specifies sequences of state transitions. This enables transparent, reproducible answers to clinically meaningful questions, such as “What is the probability that a 10-year-old who self-harms will attempt suicide within the next three years?” or “How does escalation risk differ between passive and active ideation?” Compared to case-based enumeration, this approach is auditable and scalable. Critically, every queried trajectory yields, in addition to a probability estimate, a measure of uncertainty via bootstrapped confidence intervals and entropy. In this way, the model can support more nuanced characterisation of longitudinal risk patterns and inform the development of monitoring and prevention strategies. The framework may also be useful for identifying moderators and mediators of psychiatric risk in future longitudinal or trial-based studies.

**Limitations**

*Generalizability and attrition bias in the ABCD data*

Although the ABCD Study was designed to approximate the sociodemographic composition of U. S. children,[12] prior analyses have shown sampling bias toward families with married, working parents and higher income levels.[45] Longitudinal missingness and attrition could compound these effects. While formal withdrawal from the study stayed rare, missed visits were associated with, among others, being African American, lower family income and lower parental education level.[46] These may limit the generalizability of our estimates through under-representation of families experiencing structural disadvantages.

*Markov property assumption*

The framework assumes the Markov property where a transition depends only on the current state and not on prior history. This assumption was not formally tested, and transition probabilities are likely influenced by trajectory history beyond the most recent observation. However, the framework was specifically designed to characterize and query a dataset with a few time points and rare events of interest, and first-order Markov models are widely used as pragmatic approximations in such cases when higher-order models would be too data-hungry and harder to interpret. This limitation is also partially addressed by the multi-year trajectory queries and first-passage probabilities implemented here, which can capture cumulative history. The clinically coherent patterns revealed by the framework support its utility as a first-order approximation. Extension to higher-order Markov chains would be a natural next step, though it would require more data points than are currently available in the ABCD data.

*Temporal ambiguity in the ABCD reporting of NSSI and suicidal ideation and behaviour*

The KSADS-5 questionnaires used in the ABCD study assess both current (i.e., "in the past two weeks") and lifetime (i.e., "in the past" or "ever") suicidal ideation and behaviour. The phrasing introduces temporal ambiguity, as "lifetime" includes the "past two weeks." This overlap complicates the interpretation of timing, especially in follow-up assessments. A lifetime event could have occurred prior to baseline, at baseline, or between the previous and current follow-up. Even with an assumption of a mutually exclusive split, the data's annual resolution means that nearly one year is condensed into the "past" category, making it impossible to determine whether symptoms have newly emerged, persisted, or resolved since the prior report. This ambiguity is reflected in the data. For example, current or lifetime self-harm report was 714 at baseline and decreased to 533 at 1-year follow-up, and many participants endorsing a symptom at one time point would report no lifetime symptom at a follow-up. It may be informative to assess whether parent reports are more reliable across waves.

The present framework characterises group-level transition dynamics in longitudinal cohort data rather than serving as an individual-level clinical risk prediction tool; the latter use would require prospective validation in clinical samples.

**Future works**

The mechanisms linking NSSI to suicidal behaviour, including the role of transition pathways and the potential lethality of suicide attempts, warrant further investigation. How these trajectories are moderated by comorbid psychopathology, gender, ethnicity, and socioeconomic status will be important for improving identification of high-risk subgroups. The concept of transient or phase-based self-harm also warrants deeper exploration for this age group. Future studies using linkage of community-based cohorts and electronic health record could help clarify real-world trajectories and evaluate how clinical and community responses shape recovery dynamics, complementing what can be inferred from research cohorts such as the ABCD study.

Methodologically, the Markov framework presented here is readily extensible. Incorporating demographic, clinical, and environmental covariates as stratification variables would allow direct testing of moderation hypotheses within the same probabilistic structure. Applying the approach to cohorts with higher temporal resolution, different age ranges, or non-US populations would strengthen generalizability. Integrating biological measures, such as neuroimaging or genetic risk indicators, as state-defining or moderating variables is another natural extension, connecting behavioural trajectories to underlying biological mechanisms.

More broadly, the explicit state structure and probabilistic architecture of the framework make it compatible with richer temporal modelling approaches, offering a foundation for future work that integrates time-varying covariates and adaptive prediction.

## Methods

### ABCD non-suicidal self-injury and suicidal ideation and behaviour data

The ABCD Study is a large, community-ascertained cohort of U.S. children recruited at age 9–10 through probability-based school sampling within 21 research sites[12], designed to approximate U.S. demographic distributions, though not a strict epidemiological population sample. Baseline data were collected in 2016–2018 with annual follow-ups thereafter.[12,26] In the present study, we analysed data from 11,864 children with child-report KSADS data at baseline and at ages 10–11, 11–12, and 12–13.

Details about the ABCD assessments of non-suicidal self-injury (NSSI) and suicidal ideation and behaviour were obtained from the ABCD Study publication,[26] Release Note 5.1., and examination of the data. Assessments included annual child self-report and biannual parent report using the youth version of K-SADS-5.[26,47] At each assessment, the participant answered 39 questionnaires (Supplementary Table 1) on the current ("in the past two weeks") and lifetime ("in the past" or "ever") NSSI and suicidal ideation and behaviour of the child. Based on the responses, the ABCD team derived binary (presence or absence) DSM-5 diagnostic classifications. These variables were provided separately for child and parent reports, and for current and lifetime indicators. In this study, we used the binary DSM-5 diagnostic classification from ABCD Release 5.1 (July 2023), which contained complete data from baseline (age 9-10) to the 3-year follow-up (age 12-13) and partial data from the 4-year follow up (age 13-14) available at the time of release. Child and parent data were analysed separately but current and lifetime reports were aggregated to account for the temporal overlap and to capture any emergence of symptoms within the full year-to-year interval. The binary classifications were grouped into five groups: i) NSSI, ii) passive suicidal ideation, iii) active suicidal ideation, iv) suicidal preparatory action, and v) suicidal attempt, as tabulated in Table 4.

**Data encoding and state definition**

Many children had more than one report (e.g. reporting both NSSI and active ideation). To account for all possible responses, each combination of reports was binary-encoded and converted into a decimal composite score as shown in Extended Data Fig. 1. The minimum composite score was 0 (no suicidal ideation or behaviours) and the maximum score was 31 (NSSI and all suicidal ideation and behaviour reported). The scores were ordered from passive suicidal ideation to suicidal attempt. The score parity indicated the presence (odd) or absence (even) of co-reported NSSI.

Along with the exhaustive tabulation, we summarized each participant's data by i) their most advanced reported state (passive ideation, active ideation, behaviour) and ii) presence or absence of NSSI. Suicidal preparatory action and suicidal attempt were grouped into "suicidal behaviour." This yielded eight states:

1. No NSSI or suicidal ideation/behaviour
2. NSSI only
3. Passive ideation without NSSI
4. Passive ideation with NSSI
5. Active ideation without NSSI
6. Active ideation with NSSI
7. Suicidal behaviour without NSSI
8. Suicidal behaviour with NSSI

**Descriptive analysis**

Descriptive statistics for prevalence, co-occurrence, and child-parent reporting discrepancy were computed for each year. Prevalence was calculated both per report and per participant based on the participant's most advanced state of suicidal ideation or behaviour. Co-occurrence was described as percentages of children reporting one type of symptom (e.g. NSSI) reporting another (e.g. suicidal ideation) at the same time. Discrepancies between the child- and parent-reports at each overlapping time point (baseline, 2-year follow-up (age 11-12), 4-year follow-up (age 13-14)) were also calculated. A discrepancy was defined as a report made by one informant but not the other.

**Probabilistic modelling of longitudinal transitions**

Exploratory visualization of year-to-year trajectories over the four-year period using frequency-weighted line plots (Extended Data Figure 2) using the states defined above revealed substantial within-individual variability and fluctuation across time points. We also noted that this score was useful but not a true metric. The score combined overlapping but partially distinct phenomena (NSSI, suicidal ideation, suicidal behaviour), and a change in the score reflected a transition between qualitatively distinct states, rather than a linear increase or decrease. The numerical distance between adjacent values was also not consistent in magnitude or meaning. As such, the measure violated key metric properties such as symmetry and the triangle inequality, making it unsuitable for distance-based modelling approaches. Given this pseudo-metric structure, the limited number of time points, and high volatility of individual trajectories, complex nonlinear or clustering-based models were difficult to apply. These, and the clinical relevance of the states themselves, motivated us to use a Markov approach, focusing on state-based transitions.

*Time-inhomogeneous discrete-time Markov chain*

A Markov chain is a mathematical framework for modelling a sequence of transitions between a finite set of states, where each transition is governed by a probability. These probabilities can be estimated directly from observed data and summarized in a transition matrix $P$, where each element represents the probability of moving from one state to another in the next time step. In this study, $P$ is constructed from year-to-year transitions (Figure 2), with rows representing the current state at time $t$ and columns representing the subsequent state at time $t + 1$. Mathematically, the transition probability from state $i$ at time $t$ to state $j$ at time $t + 1$ is defined as:

$$P_{t \to t+1}(i, j) = \Pr(X_{t+1} = j \mid X_t = i).$$

Because the data were collected at fixed intervals, this constitutes a discrete-time Markov chain (DTMC).

Unlike standard Markov chain applications that assume time-homogenous transitions, the state-to-state transition probabilities in our data are likely age-dependent and vary over time.

For example, the probability of passive suicidal ideation progressing to suicidal behaviour between ages 9-10 may differ from the same transition between ages 12-13. To account for this, we constructed a transition matrix for each time point, as $P_{BL \to Y1}, P_{Y1 \to Y2}$, and so on. Because the time points are discrete and the transition probabilities are time-varying, the resulting model constitutes a discrete-time, time-inhomogeneous Markov chain.[48]

Figure 2 visualizes the resulting year-to-year transition probability matrices. Given the full annual data from four time points (age 9 to13) we constructed three year-to-year transition probability matrices (Figure 2). In each matrix, the value in the $i^{th}$ row and $j^{th}$ column represents the probability of a child reporting $i^{th}$ state in one year reporting $j^{th}$ state in the following year. For example, the mean probability of a child reporting active suicidal ideation without NSSI at baseline (first matrix, row 5) transitioning to suicidal behaviour without NSSI within a year (first matrix, column 7) is about 1%. Each $i^{th}$ row contains all probabilities of a $i^{th}$ state transitioning to any state in the following year. Probabilities in a row sum to one, but due to sampling, the plotted mean values may differ by small amounts. The diagonal cells are the probabilities of staying in the same state. The upper-left region represents moving up in the pseudo-scale of the states, and the lower-right region represents moving down.

*Multi-year trajectories and first-passage probabilities*

With the year-to-year transition matrices constructed, we can compute the probability of transitioning between any two states across consecutive years using the Chapman-Kolmogorov equation:

$$P_{t \to t+n} = P_{t \to t+1} \cdot P_{t+1 \to t+2} \cdot \ldots \cdot P_{t+n-1 \to t+n}$$

This allows us to estimate the probability of ending at a given state after $\boldsymbol{n}$ years, regardless of the path taken.

In some cases, it may be informative to know the probability of reaching a critical state (e.g. suicidal attempt) for the first time within a time window. For example, a researcher may wish to estimate the probability that a child with baseline passive ideation progresses to attempt within the next 3 years. This can be addressed using the first-passage probability $\boldsymbol{F_{t \to t+x}(m, n)}$, defined as the probability of first reaching a state $\boldsymbol{S_n}$ at time $\boldsymbol{t + x}$ given that the individual started in state $\boldsymbol{S_m}$ at time $\boldsymbol{t}$.[49] The cumulative probability of transitioning to the target state at in any point within the window $(\boldsymbol{t \to t + x})$ is then given by the sum $\sum_{i=1}^{x} \boldsymbol{F_{t \to t+i}(m, n)}$.

*Bootstrapping*

When computed on the full dataset, Markov chain transition probability estimates can be sensitive to sampling noise, particularly for low-frequency transitions. To quantify sampling uncertainty and assess the stability of these estimates, we implemented a bootstrapping procedure. For each time point $\boldsymbol{t}$ and each state pair $(\boldsymbol{i, j})$, the transition probability $\boldsymbol{P_t(i, j)}$ was estimated across $\boldsymbol{k = 100}$ bootstrap resamples, each constructed by sampling $\boldsymbol{N/2}$ participants with replacement. The resulting distribution of $\boldsymbol{P_{t_k}(i, j)}$ reflects sampling variability. Its standard deviation was used as a summary measure of sampling uncertainty, and 95% confidence intervals were computed from the bootstrap distribution using t-based intervals around the mean.

**Longitudinal transition uncertainty using entropy**

To formally characterise the within-individual variability over time visualized in Extended Data Figure 2, we quantified the uncertainty in year-to-year transitions using normalised Shannon entropy,[24] which measures how predictable or diffuse the future state distribution is from a given state. For a state $\boldsymbol{i}$ at time $\boldsymbol{t}$, the entropy $\boldsymbol{E_t[i]}$ is defined as

$$\boldsymbol{E_t[i]} = -\frac{\sum_{j=1}^{N} \boldsymbol{P_t[i, j] \log_2 P_t[i, j]}}{\boldsymbol{\log_2 N}}$$

where $\boldsymbol{P_t}[\boldsymbol{i}, \boldsymbol{j}]$ is the probability of transitioning from state $\boldsymbol{i}$ to state $\boldsymbol{j}$, and $\boldsymbol{N}$ is the total number of states. Whereas bootstrapping allows for quantification of sampling uncertainty in the estimation of transition probabilities, Shannon entropy quantifies the inherent uncertainty of the system's dynamics – the unpredictability of future states given the current one – based on how evenly probability mass is distributed across possible next states.

An entropy of $\boldsymbol{E_t}[\boldsymbol{j}] = \boldsymbol{0}$ indicates a fully deterministic transition and $\boldsymbol{E_t}[\boldsymbol{j}] = \boldsymbol{1}$ reflects maximum uncertainty, where all next states are equally likely. Low entropy implies high predictability, and high entropy signals unpredictability in the individual's diagnostic course, supporting the need for more frequent or intensive follow-up.

**Custom-defined trajectories and statistical comparison**

Building on the transition matrices, multi-year and first-passage probabilities, and bootstrapping, we implemented probability and uncertainty estimation of custom-defined trajectories for any longitudinal pattern of interest using a regular expression syntax that matches the specific sequence of state transitions. For example, the pattern "^1*[7,8]" captures all possible trajectories that begins in State 1 (no report) at baseline and reaches State 7 or 8 (suicidal behaviour with or without NSSI) at any later time point.

Two or more trajectories can be statistically compared based on their bootstrapped probability distributions using tests such as Kruskal-Wallis test, reporting mean differences, approximate risk ratio, 95% confidence intervals, and p-values. Results can be visualized graphically or exported in tabular format for further analysis. This system enables flexible and interpretable testing of hypotheses on longitudinal risk, built directly on the Markov framework.

**Use of large language models (LLMs)**

Large language models (OpenAI ChatGPT and Anthropic Claude) were used during the preparation of this manuscript to assist with drafting and editing text and with software development for the MarkovPatterns.jl package. All LLM-generated outputs were reviewed, verified, and edited by the authors, who take full responsibility for the content of this manuscript.

## Acknowledgments

This work was supported by the Digital Youth programme, funded by UK Research and Innovation, and by the Nottingham Biomedical Research Centre, funded by the National Institute for Health and Care Research. The ABCD Study is supported by the National Institutes of Health and additional federal partners. The views expressed are those of the authors and not necessarily those of the funders.

## Author contributions

S.L. conceived the study, developed the modelling framework with B.C., performed the analyses, interpreted the results, and drafted the manuscript. B.C. contributed to methodological implementation and software development. M.E., N.J., E.T., K.S., P.F., Aja M., J.L., Ayan M., C.H., R.O., and D.A. contributed intellectual input on the study, interpretation of the findings, and critical revision of the manuscript. D.A. and R.O. supervised the work. All authors reviewed and approved the final manuscript.

## Competing interests

The authors declare no competing interests.

## Data availability

ABCD Study data is available through the ABCD Study Team: https://abcdstudy.org/scientists/data-sharing/

## Code availability

Source code is available at https://github.com/slee04/MarkovPatterns.jl. The project is implemented in Julia (AGPLv3).

## References

1. Brager-Larsen, A., Zeiner, P., Klungsøyr, O. & Mehlum, L. Is age of self-harm onset associated with increased frequency of non-suicidal self-injury and suicide attempts in adolescent outpatients? *BMC Psychiatry* **22**, 1–9 (2022).
2. Hink, A. B., Killings, X., Bhatt, A., Ridings, L. E. & Andrews, A. L. Adolescent Suicide—Understanding Unique Risks and Opportunities for Trauma Centers to Recognize, Intervene, and Prevent a Leading Cause of Death. *Curr. Trauma Rep.* **8**, 41 (2022).
3. Moran, P. *et al.* The Lancet Commission on self-harm. *The Lancet* **404**, 1445–1492 (2024).
4. Liu, R. T., Walsh, R. F. L., Sheehan, A. E., Cheek, S. M. & Sanzari, C. M. Prevalence and Correlates of Suicide and Nonsuicidal Self-injury in Children: A Systematic Review and Meta-analysis. *JAMA Psychiatry* **79**, 718 (2022).
5. Klonsky, E. D., May, A. M. & Saffer, B. Y. Suicide, Suicide Attempts, and Suicidal Ideation. *Annu. Rev. Clin. Psychol.* **12**, 307–330 (2016).
6. Grandclerc, S., De Labrouhe, D., Spodenkiewicz, M., Lachal, J. & Moro, M. R. Relations between Nonsuicidal Self-Injury and Suicidal Behavior in Adolescence: A Systematic Review. *PLoS ONE* **11**, e0153760 (2016).
7. Szewczuk-Bogusławska, M., Kowalski, K., Bogudzińska, B. & Misiak, B. Are the functions of non-suicidal self-injury associated with its persistence and suicide risk in university students? Insights from a network analysis. *Front. Psychiatry* **15**, 1442930 (2024).

8. Shi, T. *et al.* The Relationships Between Nonsuicidal Self-Injury, Connectedness, and Suicide Risk in Youth Presenting to the Emergency Department. *JAACAP Open* https://doi.org/10.1016/J.JAACOP.2025.01.001 (2025) doi:10.1016/J.JAACOP.2025.01.001.
9. Cipriano, A., Cella, S. & Cotrufo, P. Nonsuicidal self-injury: A systematic review. *Front. Psychol.* **8**, 282818 (2017).
10. Griep, S. K. & MacKinnon, D. F. Does Nonsuicidal Self-Injury Predict Later Suicidal Attempts? A Review of Studies. *Arch. Suicide Res.* **26**, 428–446 (2022).
11. Ammerman, B. A., Burke, T. A., O'Loughlin, C. M. & Hammond, R. The association between nonsuicidal and suicidal self-injurious behaviors: A systematic review and expanded conceptual model. *Dev. Psychopathol.* 1–16 (2025) doi:10.1017/S095457942500001X.
12. Garavan, H. *et al.* Recruiting the ABCD sample: Design considerations and procedures. *Dev. Cogn. Neurosci.* **32**, 16 (2018).
13. Wallace, G. T. & Conner, B. T. Longitudinal panel networks of risk and protective factors for early adolescent suicidality in the ABCD sample. *Dev. Psychopathol.* 1–17 (2024) doi:10.1017/S0954579424001597.
14. Wångby-Lundh, M., Lundh, L. G., Claréus, B., Bjärehed, J. & Daukantaitė, D. Developmental pathways of repetitive non-suicidal self-injury: predictors in adolescence and psychological outcomes in young adulthood. *Child Adolesc. Psychiatry Ment. Health* **17**, (2023).
15. Uno, A. *et al.* Suicidal Thoughts and Trajectories of Psychopathological and Behavioral Symptoms in Adolescence. *JAMA Netw. Open* **7**, e2353166–e2353166 (2024).

16. Gariepy, G. *et al.* Dynamic Simulation Models of Suicide and Suicide-Related Behaviors: Systematic Review. *JMIR Public Health Surveill.* **10**, e63195 (2024).

17. Townsend, E. *et al.* Uncovering key patterns in self-harm in adolescents: Sequence analysis using the Card Sort Task for Self-harm (CaTS). *J. Affect. Disord.* **206**, 161–168 (2016).

18. Wadman, R. *et al.* A sequence analysis of patterns in self-harm in young people with and without experience of being looked after in care. *Br. J. Clin. Psychol.* **56**, 388–407 (2017).

19. Lockwood, J. *et al.* A comparison of temporal pathways to self-harm in young people compared to adults: A pilot test of the Card Sort Task for Self-harm online using Indicator Wave Analysis. *Front. Psychiatry* **13**, 938003 (2023).

20. Oppenheimer, C. W., Glenn, C. R. & Miller, A. B. Future Directions in Suicide and Self-Injury Revisited: Integrating a Developmental Psychopathology Perspective. *J. Clin. Child Adolesc. Psychol.* **51**, 242–260 (2022).

21. Lewis, C. P., Klimes-Dougan, B., Croarkin, P. E. & Cullen, K. R. Understanding the emergence of suicidal thoughts and behaviors in adolescence from a brain and behavioral developmental perspective. *Neuropsychopharmacology* **51**, 259–272 (2026) doi:10.1038/s41386-025-02168-2.

22. López, R., Turnamian, M. R. & Liu, R. T. Prospective Relations between Life Stress, Emotional Clarity, and Suicidal Ideation in an Adolescent Clinical Sample. *J. Clin. Child Adolesc. Psychol.* **53**, 944–957 (2024).

23. King, C. A. & Merchant, C. R. Social and Interpersonal Factors Relating to Adolescent Suicidality: A Review of the Literature. *Arch. Suicide Res. Off. J. Int. Acad. Suicide Res.* **12**, 181–196 (2008).

24. Shannon, C. E. A Mathematical Theory of Communication. *Bell Syst. Tech. J.* **27**, 379–423 (1948).

25. Esposito, C., Dragone, M., Affuso, G., Amodeo, A. L. & Bacchini, D. Prevalence of engagement and frequency of non-suicidal self-injury behaviors in adolescence: an investigation of the longitudinal course and the role of temperamental effortful control. *Eur. Child Adolesc. Psychiatry* **32**, 2399–2414 (2023).

26. Barch, D. M. *et al.* Demographic, physical and mental health assessments in the adolescent brain and cognitive development study: Rationale and description. *Dev. Cogn. Neurosci.* **32**, 55–66 (2018).

27. Ribeiro, J. D. *et al.* Self-injurious thoughts and behaviors as risk factors for future suicide ideation, attempts, and death: a meta-analysis of longitudinal studies. *Psychol. Med.* **46**, 225–236 (2016).

28. Mars, B. *et al.* Predictors of future suicide attempt among adolescents with suicidal thoughts or non-suicidal self-harm: a population-based birth cohort study. *Lancet Psychiatry* **6**, 327–337 (2019).

29. Hamza, C. A., Stewart, S. L. & Willoughby, T. Examining the link between nonsuicidal self-injury and suicidal behavior: A review of the literature and an integrated model. *Clin. Psychol. Rev.* **32**, 482–495 (2012).

30. Andover, M. S., Morris, B. W., Wren, A. & Bruzzese, M. E. The co-occurrence of non-suicidal self-injury and attempted suicide among adolescents: distinguishing risk factors and psychosocial correlates. *Child Adolesc. Psychiatry Ment. Health* **6**, 11 (2012).

31. Voss, C., Hoyer, J., Venz, J., Pieper, L. & Beesdo-Baum, K. Non-suicidal self-injury and its co-occurrence with suicidal behavior: An epidemiological-study among adolescents and young adults. *Acta Psychiatr. Scand.* **142**, 496–508 (2020).

32. Van Orden, K. A. *et al.* The Interpersonal Theory of Suicide. *Psychol. Rev.* **117**, 575–600 (2010).

33. O'Connor, R. C. & Kirtley, O. J. The integrated motivational-volitional model of suicidal behaviour. *Philos. Trans. R. Soc. B Biol. Sci.* **373**, (2018).

34. Joiner, T. E. *et al.* Main Predictions of the Interpersonal-Psychological Theory of Suicidal Behavior: Empirical Tests in Two Samples of Young Adults. *J. Abnorm. Psychol.* **118**, 634–646 (2009).

35. Chapman, A. L., Gratz, K. L. & Brown, M. Z. Solving the puzzle of deliberate self-harm: The experiential avoidance model. *Behav. Res. Ther.* **44**, 371–394 (2006).

36. Liu, R. T. Characterizing the course of non-suicidal self-injury: A cognitive neuroscience perspective. *Neurosci. Biobehav. Rev.* **80**, 159–165 (2017).

37. Koenig, J. *et al.* High-frequency ecological momentary assessment of emotional and interpersonal states preceding and following self-injury in female adolescents. *Eur. Child Adolesc. Psychiatry* **30**, 1299–1308 (2021).

38. Franklin, J. C. *et al.* Risk factors for suicidal thoughts and behaviors: A meta-analysis of 50 years of research. *Psychol. Bull.* **143**, 187–232 (2017).

39. Sherman, L. E. *et al.* Development of the Default Mode and Central Executive Networks across early adolescence: A longitudinal study. *Dev. Cogn. Neurosci.* **10**, 148–159 (2014).

40. Borbás, R. *et al.* Evolving brain function and connectivity patterns during mentalizing in children and adults. *Commun. Biol.* **9**, 282 (2026).

41. Crone, E. A. & van Drunen, L. Development of Self-Concept in Childhood and Adolescence: How Neuroscience Can Inform Theory and Vice Versa. *Hum. Dev.* **68**, 255–271 (2024).

42. Casey, B. J., Jones, R. M. & Hare, T. A. The Adolescent Brain. *Ann. N. Y. Acad. Sci.* **1124**, 111–126 (2008).

43. Somerville, L. H. The Teenage Brain: Sensitivity to Social Evaluation. *Curr. Dir. Psychol. Sci.* **22**, 121–127 (2013).

44. Foulkes, L. & Blakemore, S.-J. Is there heightened sensitivity to social reward in adolescence? *Curr. Opin. Neurobiol.* **40**, 81–85 (2016).

45. Heeringa, S. G. & Berglund, P. A. A Guide for Population-based Analysis of the Adolescent Brain Cognitive Development (ABCD) Study Baseline Data. *bioRxiv* 2020.02.10.942011 (2020) doi:10.1101/2020.02.10.942011.

46. Feldstein Ewing, S. W. *et al.* Measuring retention within the adolescent brain cognitive development (ABCD)SM study. *Dev. Cogn. Neurosci.* **54**, 101081 (2022).

47. Kaufman, J. *et al.* Schedule for Affective Disorders and Schizophrenia for School-Age Children-Present and Lifetime Version (K-SADS-PL): initial reliability and validity data. *J. Am. Acad. Child Adolesc. Psychiatry* **36**, 980–988 (1997).

48. Ching, W. K., Ng, M. K. & Fung, E. S. Higher-order multivariate Markov chains and their applications. *Linear Algebra Its Appl.* **428**, 492–507 (2008).

49. Norris, J. R. *Markov Chains*. (Cambridge University Press, 1997).

**Figure 1.**

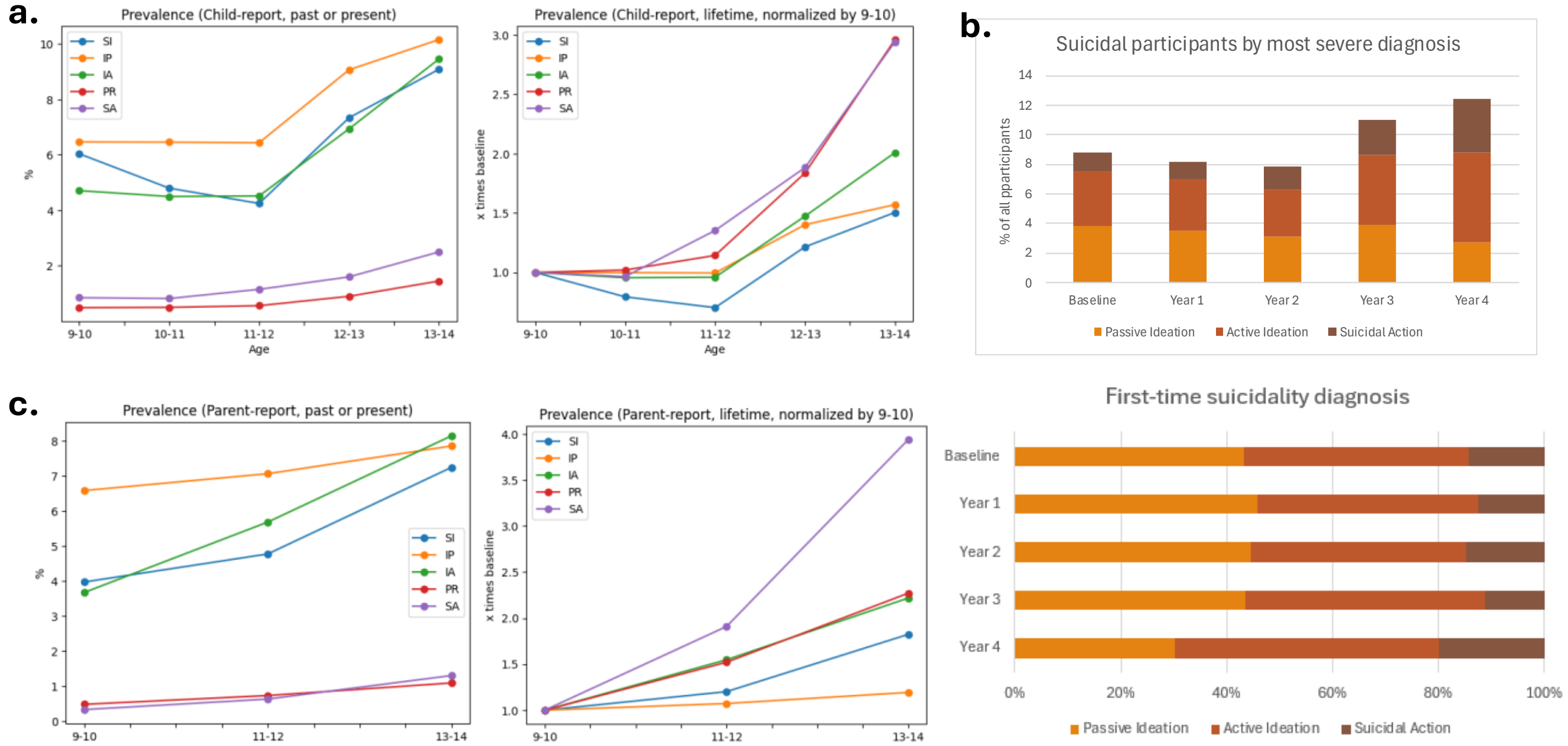

Child-report (a) and parent-report (b) prevalence of NSSI and suicidality in the ABCD cohort. Left: percentage; Right: percentage normalised by baseline. SI: self-injury; IP: passive ideation; IA: active ideation; PR: preparatory action; SA: suicidal attempt. b. Parent-report prevalence of self-harm and suicidality in the ABCD cohort. C. Percentage of suicidal participants by most severe diagnosis

**Figure 2**

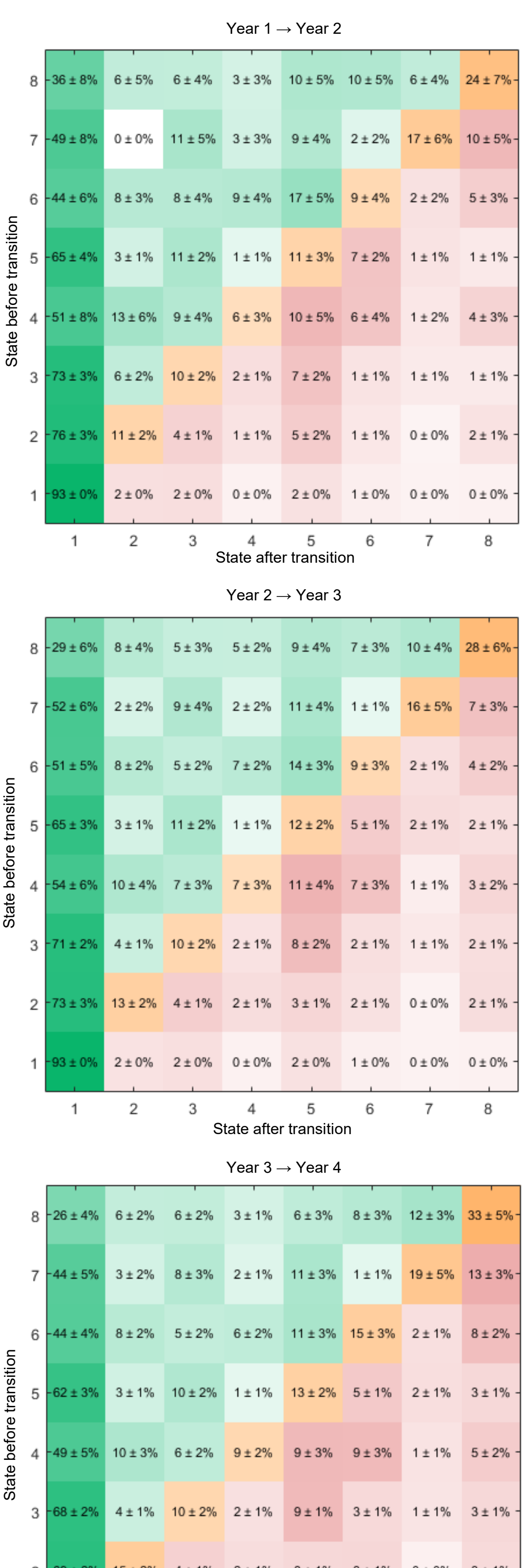

Year-to-year suicidality and NSSI transition probability matrices based on eight states. In each matrix, the states are defined as: 1 = no suicidality or nonsuicidal self-injury (NSSI); 2 = NSSI; 3 = passive suicidal ideation without NSSI; 4 = passive suicidal ideation with NSSI; 5 = active suicidal ideation without NSSI; 6 = active suicidal ideation with NSSI; 7 = suicidal action without NSSI; 8 = suicidal action with NSSI.
Top: From Baseline (age 9-10) to Year 1 follow-up (age 10-11). Middle: From Year 1 (age 10-11) to Year 2 follow-up (age 11-12). Bottom: From Year 2 (age 11-12) to Year 3 follow-up (age 12-13).

## Figure 3

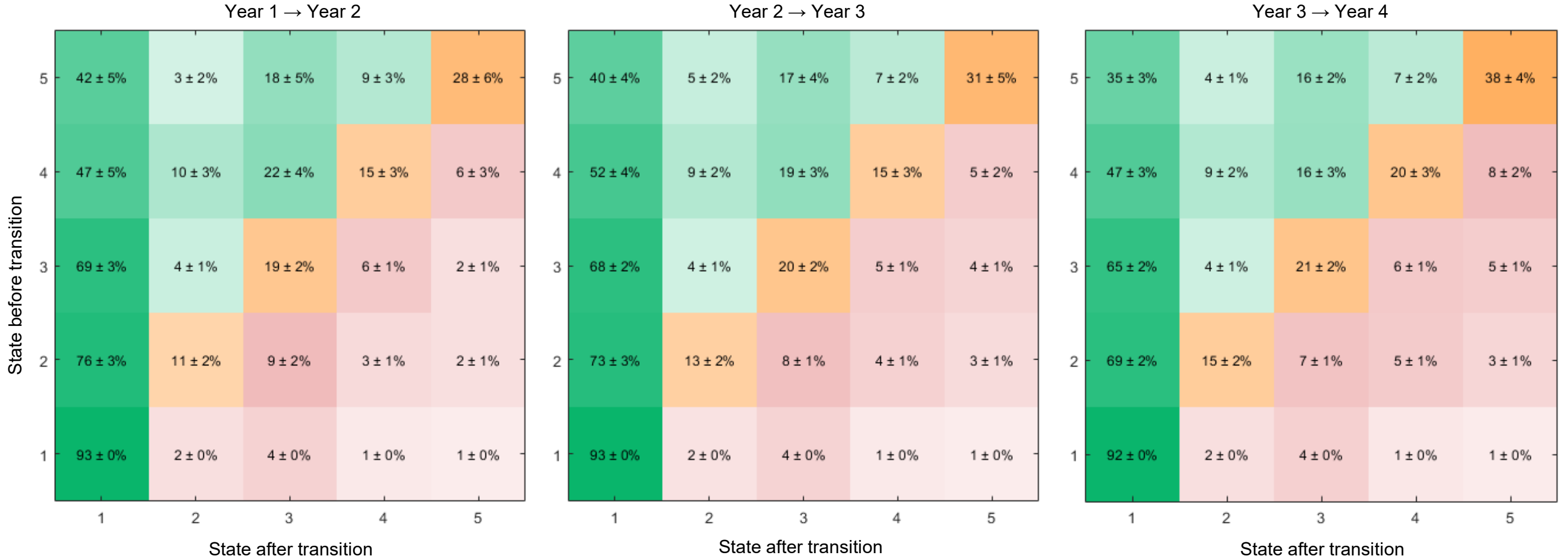

Year-to-year suicidality and NSSI transition probability matrices based on five (composite) states. In each matrix, the states are defined as: 1 = no suicidality or nonsuicidal self-injury (NSSI); 2 = NSSI; 3 = suicidal ideation without NSSI (active or passive); 4 = suicidal ideation with NSSI (active or passive); 5 = suicidal action, with or without NSSI.
Left: From Baseline (age 9-10) to Year 1 follow-up (age 10-11). Middle: From Year 1 (age 10-11) to Year 2 follow-up (age 11-12). Right: From Year 2 (age 11-12) to Year 3 follow-up (age 12-13).

**Figure 4**

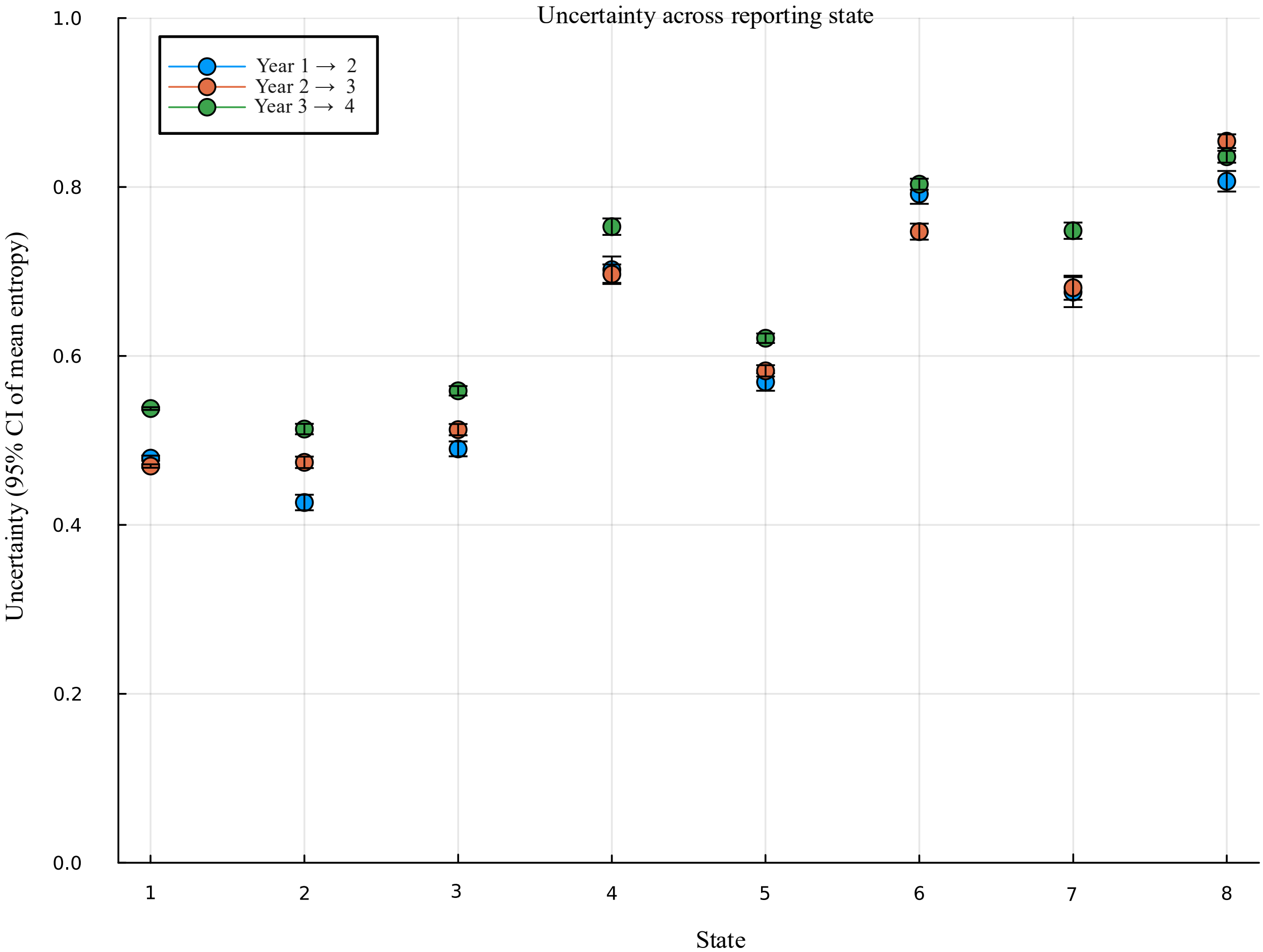

Mean Shannon entropy of year-to-year transitions based on the starting state. For example, the blue circle in State 3 represents the uncertainty in predicting the 1-year follow-up state from passive ideation (without NSSI) at Baseline. The solid lines pair the with- and without-NSSI states, showing that suicidality with NSSI (State 4, 6, 8) are less predictable than suicidality without NSSI (State 3, 5, 7).

**Figure 5**

**a.**

**b.**

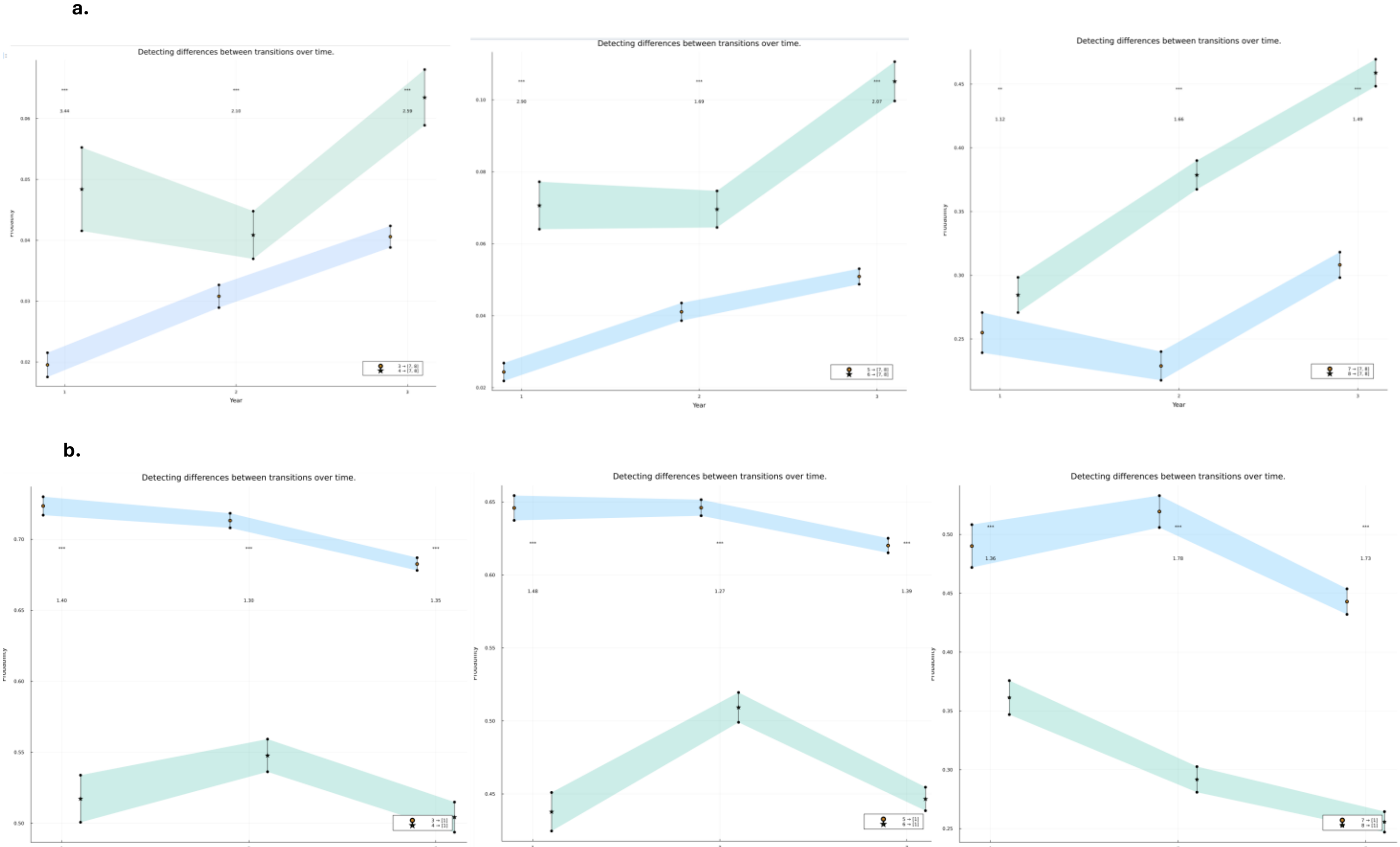

a. Comparison of year-to-year transition probabilities to suicidal action (State 7 or 8), between passive ideation without NSSI (State 3) and with NSSI (State 4), active ideation without NSSI (State 5) and with NSSI (State 6), and suicidal action without NSSI (State 7) and with NSSI (State 8). b. Comparison of year-to-year transition probabilities of recovering to no report (State 1) between passive ideation without NSSI (State 3) and with NSSI (State 4), active ideation without NSSI (State 5) and with NSSI (State 6), and suicidal action without NSSI (State 7) and with NSSI (State 8).

**Figure 6**

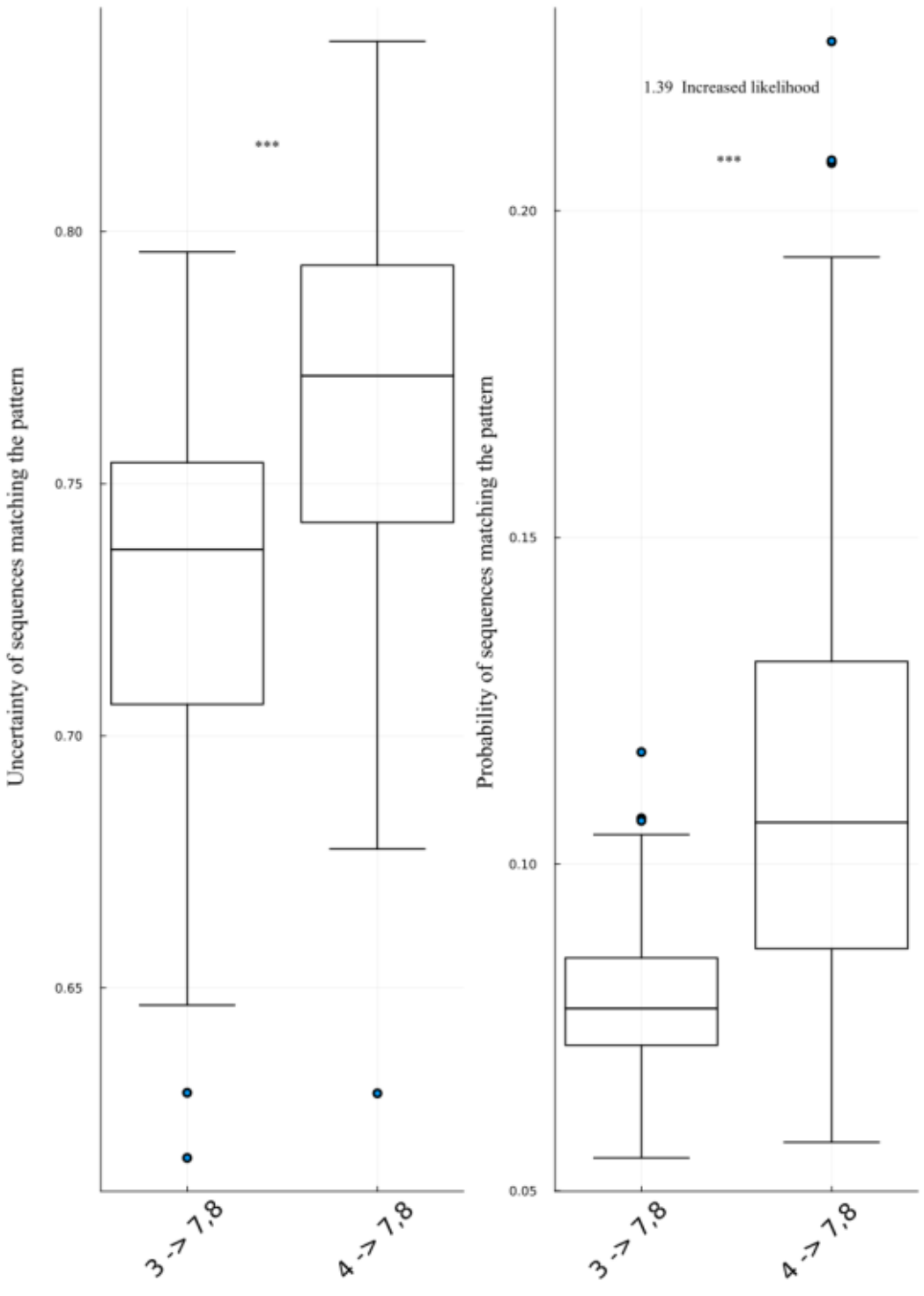

Comparison of 3-year transition probabilities to suicidal action between passive ideation without and with NSSI.

*Table 1.* Prevalence of non-suicidal self-injury (NSSI) and suicidal ideation / action in the ABCD cohort

a. Based on child-report

| Time point | Age (years) | Total | NSSI | | Passive Ideation | | Active Ideation | | Prep Action | | Attempt | |
|---|---|---|---|---|---|---|---|---|---|---|---|---|
| | | N | n | % | N | % | n | % | n | % | N | % |
| Baseline | 9-10 | 11,812 | 714 | 6.04 | 764 | 6.47 | 556 | 4.71 | 58 | 0.49 | 100 | 0.85 |
| Year 1 | 10-11 | 11,095 | 533 | 4.80 | 717 | 6.46 | 499 | 4.50 | 56 | 0.50 | 91 | 0.82 |
| Year 2 | 11-12 | 10,883 | 462 | 4.25 | 701 | 6.44 | 492 | 4.52 | 61 | 0.56 | 125 | 1.15 |
| Year 3 | 12-13 | 10,326 | 758 | 7.34 | 937 | 9.07 | 717 | 6.94 | 93 | 0.90 | 165 | 1.60 |
| Year 4 | 13-14 | 4,674 | 425 | 9.09 | 475 | 10.2 | 442 | 9.46 | 68 | 1.45 | 117 | 2.50 |

b. Based on parent-report

| Time point | Age | Total | NSSI | | Passive Ideation | | Active Ideation | | Prep Action | | Attempt | |
|---|---|---|---|---|---|---|---|---|---|---|---|---|
| | | n | n | % | n | % | n | % | n | % | N | % |
| Baseline | 9-10 | 11,747 | 466 | 3.97 | 773 | 6.58 | 431 | 3.67 | 56 | 0.48 | 39 | 0.33 |
| Year 2 | 11-12 | 10,756 | 513 | 4.77 | 759 | 7.06 | 611 | 5.68 | 79 | 0.73 | 68 | 0.63 |
| Year 4 | 13-14 | 4,754 | 344 | 7.24 | 373 | 7.85 | 387 | 8.14 | 52 | 1.09 | 62 | 1.30 |

Non-suicidal self-injury, passive suicidal ideation, active suicidal ideation, preparatory action for suicide, and suicide attempt were defined based on DSM-V diagnoses and grouping as described in Methods.

*Table 2.* Co-report of NSSI and suicidal ideation or behaviour in the ABCD cohort

a. NSSI and suicidal ideation or behaviour

| | Baseline | Year 1 | Year 2 | Year 3 | Year 4 |
|---|---|---|---|---|---|
| NSSI (n) | 714 | 533 | 462 | 758 | 425 |
| NSSI with suicidality (n) | 291 | 254 | 257 | 433 | 275 |
| % of NSSI co-reporting suicidality | 40.8 | 47.7 | 55.6 | 57.1 | 64.7 |
| | | | | | |
| Suicidality (n) | 1040 | 910 | 858 | 1135 | 583 |
| Suicidality with NSSI(n) | 291 | 254 | 257 | 433 | 275 |
| % of suicidality co-reporting NSSI | 28.0 | 27.9 | 30.0 | 38.1 | 47.2 |
| | | | | | |
| % of suicidal children co-reporting NSSI by type of suicidality | | | | | |
| Passive ideation | 18.4 | 18.8 | 17.8 | 15.3 | 21.4 |
| Active ideation | 29.7 | 26.5 | 26.2 | 40.8 | 42.5 |
| Action | 52.0 | 58.2 | 60.8 | 70.8 | 74.1 |

b. Suicidal ideation, preparatory action, and attempt

| | Baseline | Year 1 | Year 2 | Year 3 | Year 4 |
|---|---|---|---|---|---|
| Active ideation as the most severe report (n) | 441 | 388 | 344 | 495 | 287 |
| % co-reporting passive ideation (%) | 48.3 | 57.7 | 63.7 | 65.3 | 65.9 |
| | | | | | |
| Suicidal attempt as the most severe report (n) | 99 | 91 | 125 | 165 | 117 |
| % co-reporting preparatory action (%) | 10.1 | 14.3 | 8.0 | 10.9 | 12.8 |
| % co-reporting suicidal ideation (%) | 84.8 | 92.3 | 89.6 | 97.0 | 95.7 |

*Table 3.* Child-parent report discrepancy in non-suicidal self-injury (NSSI) and suicidal ideation / action in the ABCD cohort

| Baseline | NSSI | Passive Ideation | Active Ideation | Prep action | Attempt |
|---|---|---|---|---|---|
| Child only (% of report) | 599 (84%) | 589 (77%) | 451 (81%) | 58 (100%) | 87 (87%) |
| Parent only (% of report) | 354 (76%) | 594 (77%) | 327 (76%) | 55 (98%) | 27 (70%) |
| Both | 110 | 174 | 102 | 1 | 12 |
| Either (% of cohort) | 1,175 (10%) | 1,357 (12%) | 880 (7.5%) | 114 (1%) | 114 (1%) |
| Year 2 | | | | | |
| Child only (% of report) | 300 (65%) | 464 (66%) | 326 (66%) | 53 (87%) | 101 (80%) |
| Parent only (% of report) | 325 (69%) | 532 (70%) | 454 (74%) | 71 (90%) | 47 (69%) |
| Both | 156 | 221 | 153 | 5 | 20 |
| Either (% of cohort) | 781 (7%) | 1,217 (11%) | 933 (8.7%) | 129 (1.2%) | 168 (1.6%) |
| Year 4 | | | | | |
| Child only (% of report) | 248 (58%) | 309 (65%) | 261 (59%) | 58 (85%) | 86 (74%) |
| Parent only (% of report) | 160 (47%) | 202 (54%) | 198 (51%) | 41 (79%) | 28 (45%) |
| Both | 177 | 166 | 181 | 10 | 31 |
| Either (% of cohort) | 585 (13%) | 677 (14%) | 640 (14%) | 109 (2.3%) | 145 (3.1%) |

Note. For the 'Child only' and 'Parent only' rows, percentages are calculated among all reports made by that reporter, including concordant reports. For the 'Either' row, percentages are calculated using the full cohort at the corresponding time point. 'Both' indicates concordant child and parent reports.

*Table 4. DSM-V descriptions of self-harm and suicidality and groups defined for the current study. The preparatory actions on the questionnaire were reported in a combination of multiple choices and written response (1=Bought or stored pills; 2=Obtained a knife; 3=Obtained a gun; 4=Obtained a rope; 5=Gave away my things; 6=Wrote a suicide note; 7=Researched methods on internet; 8=Other (type in); 0=None; I have not made any preparations).*

| **ABCD DSM-V Descriptions** | **Current Study Grouping** |
|---|---|
| Self-injurious behaviour without suicidal intent | NSSI |
| Suicidal Ideation, Passive | Passive Ideation |
| Suicidal Ideation Active, Nonspecific<br>Suicidal Ideation Active, Method<br>Suicidal Ideation Active, Intent<br>Suicidal Ideation Active, Plan | Active Ideation |
| Preparatory Actions toward imminent suicidal behaviour | Preparatory Action |
| Suicidal Attempt | Attempt |

1. ABCD KSADS-5 Questionnaires

The data was extracted from the ABCD Data Dictionary: https://data-dict.abcdstudy.org/

1.1. Child report

| Table Name | Variable Name | Variable Label | Notes |
|---|---|---|---|
| mh_y_ksads_si | ksads_suicidal_raw_1105_t | You mentioned that in the past 2 weeks you did some things to hurt yourself, like scratching, cutting, or burning yourself. Were you trying to kill yourself when you did these things? | 0=No; 1=Yes |
| mh_y_ksads_si | ksads_suicidal_raw_1106_t | Did you think that you had at least some chance that you would die as a result of what you did? | 0=No; 1=Yes |
| mh_y_ksads_si | ksads_suicidal_raw_1107_t | You mentioned in the past two weeks you thought about actually wanting to kill yourself. Have you thought about how you would do it (even if you had no intention of actually doing it)? | 0=No; 1=Yes |
| mh_y_ksads_si | ksads_suicidal_raw_1108_t | At any point in the past two weeks did you have some intention on acting on these thoughts, even if you weren't 100% sure you would do it? | 0=No; 1=Yes |
| mh_y_ksads_si | ksads_suicidal_raw_1109_t | In the past two weeks, did you think through the details of how you would do it, for instance, decide on a specific method, place, or time? | 0=No; 1=Yes |
| mh_y_ksads_si | ksads_suicidal_raw_1110_t | You mentioned earlier that in the past two weeks you had thoughts that you wished that you were dead. Have you made any preparations for killing yourself? | 1=Bought or stored pills; 2=Obtained a knife; 3=Obtained a gun; 4=Obtained a rope; 5=Gave away my things; 6=Wrote a suicide note; 7=Researched methods on internet; 8=Other (type in); 0=None; I have not made any preparations |
| mh_y_ksads_si | ksads_suicidal_raw_1111_t | Have you made any preparations for killing yourself? | 1=Bought or stored pills; 2=Obtained a knife; 3=Obtained a gun; 4=Obtained a rope; 5=Gave away my things; 6=Wrote a suicide note; |
| | | | 7=Researched methods on internet; 8=Other (type in); 0=None; I have not made any preparations |
| mh_y_ksads_si | ksads_suicidal_raw_1112_t | In the past two weeks did you start to do something to end your life, but either stopped yourself or were interrupted by someone else (for example, you were about to take pills or had a gun ready, or were about to jump or hang yourself, but either stopped yourself or were stopped by someone else)? | 1=Yes, but I stopped myself; 3=Yes, but I was stopped by someone else; 2=No, I did not start an attempt |
| mh_y_ksads_si | ksads_suicidal_raw_1113_t | You mentioned that in the past two weeks you had made a suicide attempt. What did you do? | 1=Took pills; 2=With a gun; 3=By hanging; 4=By jumping; 5=By cutting; 6=Other (type in) |
| mh_y_ksads_si | ksads_suicidal_raw_1114_t | Did you think that you had at least some chance of dying as a result? | 0=No; 1=Yes |
| mh_y_ksads_si | ksads_suicidal_raw_1115_t | You mentioned that in the past, you did some things to hurt yourself, like scratching, cutting, or burning yourself. Were you trying to kill yourself when you did these things? | 0=No; 1=Yes |
| mh_y_ksads_si | ksads_suicidal_raw_1116_t | Did you think that you had at least some chance that you would die as a result of what you did? | 0=No; 1=Yes |
| mh_y_ksads_si | ksads_suicidal_raw_1117_t | Please tell me when this happened and what you did. | |
| mh_y_ksads_si | ksads_suicidal_raw_1118_t | You mentioned in the past you thought about actually wanting to kill yourself. When was that? | |
| mh_y_ksads_si | ksads_suicidal_raw_1119_t | Did you think about how you would do it (even if you had no intention of actually doing it)? | 0=No; 1=Yes |
| mh_y_ksads_si | ksads_suicidal_raw_1120_t | Back then, at any point did you have some intention on acting on these thoughts, even if you weren't 100% sure you would do it? | 0=No; 1=Yes |
| mh_y_ksads_si | ksads_suicidal_raw_1121_t | Did you think through the details of exactly how you would do it, for instance, decide on a specific place or time? | 0=No; 1=Yes |
| mh_y_ksads_si | ksads_suicidal_raw_1122_t | Back then, did you make any preparations for killing yourself? | 0=None; I have not made any preparations; 1=Bought or stored pills; 2=Obtained a knife; 3=Obtained a gun; 4=Obtained a rope; 5=Gave away my things; 6=Wrote a suicide |
| | | | note; 7=Researched methods on internet; 8=Other (type in); 0=None; I have not made any preparations |
| mh_y_ksads_si | ksads_suicidal_raw_1123_t | You mentioned that in the past, you wished you were dead or thought you would be better off dead. Back then, did you make any preparations for killing yourself? | 1=Bought or stored pills; 2=Obtained a knife; 3=Obtained a gun; 4=Obtained a rope; 5=Gave away my things; 6=Wrote a suicide note; 7=Researched methods on internet; 8=Other (type in); 0=None; I have not made any preparations |
| mh_y_ksads_si | ksads_suicidal_raw_1124_t | Back then, did you start to do something to end your life, but either stopped yourself or were interrupted by someone else (for example, you were about to take pills or had a gun ready, or were about to jump or hang yourself, but either stopped yourself or were stopped by someone else)? | 1=Yes, but I stopped myself; 3=Yes, but I was stopped by someone else; 2=No, I did not start an attempt |
| mh_y_ksads_si | ksads_suicidal_raw_1125_t | You mentioned that there was a time in the past when you made a suicide attempt. How many times in the past have you tried to kill yourself? | |
| mh_y_ksads_si | ksads_suicidal_raw_1126_t | Please tell me when this happened and what you did. | |
| mh_y_ksads_si | ksads_suicidal_raw_1127_t | Did you think that you had at least some chance of dying as a result? | 0=No; 1=Yes |
| mh_y_ksads_si | ksads_suicidal_raw_819_t | Sometimes when kids get upset or feel numb, they may do things to hurt themselves, like scratching, cutting, or burning themselves. In the past two weeks, how often have you done any of these things or other things to try to hurt yourself? | 0=Not at all; 1=Rarely; 2=Several days; 3=More than half the days; 4=Nearly every day |
| mh_y_ksads_si | ksads_suicidal_raw_820_t | Was there ever a time in the past when you did things to hurt yourself on purpose because you were upset, like cut, scratch or burn yourself? | 0=No; 1=Yes |
| mh_y_ksads_si | ksads_suicidal_raw_821_t | Was there ever another time in the past when you did things to hurt yourself on purpose, like cut, scratch or burn yourself? | 0=No; 1=Yes |
| mh_y_ksads_si | ksads_suicidal_raw_822_t | When was that? | |
| mh_y_ksads_si | ksads_suicidal_raw_823_t | In the past two weeks, how often have you wished you were dead or had thoughts that you would be better off dead? | 0=Not at all; 1=Rarely; 2=Several days; 3=More than half the days; 4=Nearly every day |
| mh_y_ksads_si | ksads_suicidal_raw_824_t | Was there ever a time in the past when you often wished you were dead or thought you would be better off dead? | 0=No; 1=Yes |
| mh_y_ksads_si | ksads_suicidal_raw_825_t | Was there ever another time in the past when you often wished you were dead or thought you would be better off dead? | 0=No; 1=Yes |
| mh_y_ksads_si | ksads_suicidal_raw_826_t | When was that? | |
| mh_y_ksads_si | ksads_suicidal_raw_827_t | In the past two weeks, how often did you think seriously about wanting to kill yourself? | 0=Not at all; 1=Rarely; 2=Several days; 3=More than half the days; 4=Nearly every day |
| mh_y_ksads_si | ksads_suicidal_raw_828_t | Was there ever a time when you thought about wanting to kill yourself? | 0=No; 1=Yes |
| mh_y_ksads_si | ksads_suicidal_raw_829_t | Was there ever another time when you thought about wanting to kill yourself? | 0=No; 1=Yes |
| mh_y_ksads_si | ksads_suicidal_raw_830_t | In the past two weeks, did you make a suicide attempt and do something to try to kill yourself? | 0=No; 1=Yes |
| mh_y_ksads_si | ksads_suicidal_raw_831_t | I appreciate you telling me that. I will ask you more about that later. | |
| mh_y_ksads_si | ksads_suicidal_raw_832_t | Was there ever a time when you did something to try to kill yourself and actually made a suicide attempt? | 0=No; 1=Yes |
| mh_y_ksads_si | ksads_suicidal_raw_833_t | Was there ever another time when you did something to try to kill yourself and made a suicide attempt? | 0=No; 1=Yes |
| mh_y_ksads_si | ksads_suicidal_raw_834_t | I appreciate you telling me that. I will ask you more about that later. | |

*Table 1. KSADS - Suicidality (Indiv. Questions) for child*

## 1.2 Parent Report

| Table Name | Variable Name | Variable Label | Notes |
|---|---|---|---|
| mh_p_ksads_si | ksads_suicidal_raw_1105_p | You mentioned that in the past 2 weeks your child did some things to hurt himself or herself, like scratching, cutting, or | 0=No; 1=Yes |
|  |  | burning themself. Was your child trying to kill themself when he or she did these things? |  |
| mh_p_ksads_si | ksads_suicidal_raw_1106_p | Did your child think that he or she had at least some chance that they would die as a result of what they did? | 0=No; 1=Yes |
| mh_p_ksads_si | ksads_suicidal_raw_1107_p | You mentioned in the past two weeks your child thought about actually wanting to kill himself or herself. Has your child thought about how they would do it (even if they had no intention of actually doing it)? | 0=No; 1=Yes |
| mh_p_ksads_si | ksads_suicidal_raw_1108_p | At any point in the past two weeks did your child have some intention on acting on these thoughts, even if they weren't 100% sure they would do it? | 0=No; 1=Yes |
| mh_p_ksads_si | ksads_suicidal_raw_1109_p | In the past two weeks, did your child think through the details of how he or she would do it, for instance, decide on a specific method, place, or time? | 0=No; 1=Yes |
| mh_p_ksads_si | ksads_suicidal_raw_1110_p | You mentioned earlier that in the past two weeks your child had thoughts that he or she wished that they were dead. Has your child made any preparations for killing him/herself? | 1=Bought or stored pills; 2=Obtained a knife; 3=Obtained a gun; 4=Obtained a rope; 5=Gave away my things; 6=Wrote a suicide note; 7=Researched methods on internet; 8=Other (type in); 0=None; I have not made any preparations |
| mh_p_ksads_si | ksads_suicidal_raw_1111_p | Has your child made any preparations for killing him/herself? | 1=Bought or stored pills; 2=Obtained a knife; 3=Obtained a gun; 4=Obtained a rope; 5=Gave away my things; 6=Wrote a suicide note; 7=Researched methods on internet; 8=Other (type in); 0=None; I have not made any preparations |
| mh_p_ksads_si | ksads_suicidal_raw_1112_p | In the past two weeks did your child start to do something to end his or her life, but either stopped themself or were interrupted by someone else (for example, your child was about to take pills or had a gun ready, or was about to jump or | 1=Yes, but I stopped myself; 3=Yes, but I was stopped by someone else; 2=No, I did not start an attempt |

| | | hang themself, but either stopped themself or were stopped by someone else)? | |
|---|---|---|---|
| mh_p_ksads_si | ksads_suicidal_raw_1113_p | You mentioned that in the past two weeks your child had made a suicide attempt. What did your child do? | 1=Took pills; 2=With a gun; 3=By hanging; 4=By jumping; 5=By cutting; 6=Other (type in) |
| mh_p_ksads_si | ksads_suicidal_raw_1114_p | Did your child think that he or she had at least some chance of dying as a result? | 0=No; 1=Yes |
| mh_p_ksads_si | ksads_suicidal_raw_1115_p | You mentioned that in the past, your child did some things to hurt himself or herself, like scratching, cutting, or burning themself. Was your child trying to kill him or herself by doing these things? | 0=No; 1=Yes |
| mh_p_ksads_si | ksads_suicidal_raw_1116_p | Did your child think that he or she had at least some chance that they would die as a result of what they did? | 0=No; 1=Yes |
| mh_p_ksads_si | ksads_suicidal_raw_1117_p | Please tell me when this happened and what your child did. | |
| mh_p_ksads_si | ksads_suicidal_raw_1118_p | You mentioned in the past your child thought about actually wanting to kill himself or herself. When was that? | |
| mh_p_ksads_si | ksads_suicidal_raw_1119_p | Did your child think about how they would do it (even if they had no intention of actually doing it)? | 0=No; 1=Yes |
| mh_p_ksads_si | ksads_suicidal_raw_1120_p | Back then, at any point did your child have some intention on acting on these thoughts, even if they weren't 100% sure they would do it? | 0=No; 1=Yes |
| mh_p_ksads_si | ksads_suicidal_raw_1121_p | Did your child think through the details of exactly how he or she would do it, for instance, decide on a specific place or time? | 0=No; 1=Yes |
| mh_p_ksads_si | ksads_suicidal_raw_1122_p | Back then, did your child make any preparations for killing themself? | 1=Bought or stored pills; 2=Obtained a knife; 3=Obtained a gun; 4=Obtained a rope; 5=Gave away my things; 6=Wrote a suicide note; 7=Researched methods on internet; 8=Other (type in); 0=None; I have not made any preparations |
| mh_p_ksads_si | ksads_suicidal_raw_1123_p | You mentioned that in the past, your child wished he or she were dead or thought he or she would be better off dead. Back | 1=Bought or stored pills; 2=Obtained a knife; 3=Obtained a gun; |
| | | then, did he or she make any preparations for killing himself or herself? | 4=Obtained a rope; 5=Gave away my things; 6=Wrote a suicide note; 7=Researched methods on internet; 8=Other (type in); 0=None; I have not made any preparations |
| mh_p_ksads_si | ksads_suicidal_raw_1124_p | Back then, did your child start to do something to end their life, but either stopped themself or were interrupted by someone else (for example, your child was about to take pills or had a gun ready, or was about to jump or hang themself, but either stopped themself or were stopped by someone else)? | 1=Yes, but I stopped myself; 3=Yes, but I was stopped by someone else; 2=No, I did not start an attempt |
| mh_p_ksads_si | ksads_suicidal_raw_1125_p | You mentioned that were was a time in the past when your child made a suicide attempt. How many times in the past has your child attempted suicide? | |
| mh_p_ksads_si | ksads_suicidal_raw_1126_p | Please tell me when this happened and what your child did. | |
| mh_p_ksads_si | ksads_suicidal_raw_1127_p | Did your child think that they had at least some chance of dying as a result? | 0=No; 1=Yes |
| mh_p_ksads_si | ksads_suicidal_raw_819_p | Sometimes when kids get upset or feel numb, they may do some things to hurt themselves, like scratching, cutting, or burning themselves. In the past two weeks, how often has your child done any of these things or other things to try to hurt himself or herself? | 0=Not at all; 1=Rarely; 2=Several days; 3=More than half the days; 4=Nearly every day |
| mh_p_ksads_si | ksads_suicidal_raw_820_p | Was there ever a time in the past when your child did things to hurt himself or herself on purpose because your child was upset, like cut, scratch or burn himself or herself? | 0=No; 1=Yes |
| mh_p_ksads_si | ksads_suicidal_raw_821_p | Was there ever another time in the past when your child did things to hurt himself or herself on purpose, like cut, scratch or burn himself or herself? | 0=No; 1=Yes |
| mh_p_ksads_si | ksads_suicidal_raw_822_p | When was that? | |
| mh_p_ksads_si | ksads_suicidal_raw_823_p | In the past two weeks, how often has your child wished he or she was dead or had thoughts that he or she would be better off dead? | 0=Not at all; 1=Rarely; 2=Several days; 3=More than half the days; 4=Nearly every day |
| mh_p_ksads_si | ksads_suicidal_raw_824_p | Was there ever a time in the past when your child often wished he or she was dead or thought he or she would be better off dead? | 0=No; 1=Yes |
| mh_p_ksads_si | ksads_suicidal_raw_825_p | Was there ever another time in the past when your child often wished he or she was dead or thought he or she would be better off dead? | 0=No; 1=Yes |
| mh_p_ksads_si | ksads_suicidal_raw_826_p | When was that? | |
| mh_p_ksads_si | ksads_suicidal_raw_827_p | In the past two weeks, how often did your child think about actually wanting to kill himself or herself? | 0=Not at all; 1=Rarely; 2=Several days; 3=More than half the days; 4=Nearly every day |
| mh_p_ksads_si | ksads_suicidal_raw_828_p | Was there ever a time when your child thought about wanting to kill himself or herself? | 0=No; 1=Yes |
| mh_p_ksads_si | ksads_suicidal_raw_829_p | Was there ever another time when your child thought about wanting to kill himself or herself? | 0=No; 1=Yes |
| mh_p_ksads_si | ksads_suicidal_raw_830_p | In the past two weeks, did your child actually do something to kill himself or herself and make a suicide attempt? | 0=No; 1=Yes |
| mh_p_ksads_si | ksads_suicidal_raw_831_p | I appreciate you telling me that. I will ask you more about that later. | |
| mh_p_ksads_si | ksads_suicidal_raw_832_p | Was there ever a time when your child did something to try to kill himself or herself and actually made a suicide attempt? | 0=No; 1=Yes |
| mh_p_ksads_si | ksads_suicidal_raw_833_p | Was there ever another time when your child did something to try to kill himself or herself and actually made a suicide attempt? | 0=No; 1=Yes |
| mh_p_ksads_si | ksads_suicidal_raw_834_p | I appreciate you telling me that. I will ask you more about that later. | |

**Extended Data Fig. 1.** a) composite scoring based on individual suicidal ideation, suicidal behaviour, and non-suicidal self-injury reports. The colour coding indicates the most advanced suicidal ideation or behaviour reported: green is without suicidal ideation or behaviour, yellow is passive ideation, orange is active ideation, and red is suicidal behaviour. b) frequency charts of composite scores from baseline (age 9-10) to Year 4 follow up (age 13-14). As the cohort ages, there is a noticeable increase in active ideation and suicidal behaviour reports.

a)

| Attempt | Preparatory Action | Active Ideation | Passive Ideation | NSSI | **Score** |
|---|---|---|---|---|---|
| 1 | 1 | 1 | 1 | 1 | **31** |
| 1 | 1 | 1 | 1 | 0 | **30** |
| 1 | 1 | 1 | 0 | 1 | **29** |
| 1 | 1 | 1 | 0 | 0 | **28** |
| 1 | 1 | 0 | 1 | 1 | **27** |
| 1 | 1 | 0 | 1 | 0 | **26** |
| 1 | 1 | 0 | 0 | 1 | **25** |
| 1 | 1 | 0 | 0 | 0 | **24** |
| 1 | 0 | 1 | 1 | 1 | **23** |
| 1 | 0 | 1 | 1 | 0 | **22** |
| 1 | 0 | 1 | 0 | 1 | **21** |
| 1 | 0 | 1 | 0 | 0 | **20** |
| 1 | 0 | 0 | 1 | 1 | **19** |
| 1 | 0 | 0 | 1 | 0 | **18** |
| 1 | 0 | 0 | 0 | 1 | **17** |
| 1 | 0 | 0 | 0 | 0 | **16** |
| 0 | 1 | 1 | 1 | 1 | **15** |
| 0 | 1 | 1 | 1 | 0 | **14** |
| 0 | 1 | 1 | 0 | 1 | **13** |
| 0 | 1 | 1 | 0 | 0 | **12** |
| 0 | 1 | 0 | 1 | 1 | **11** |
| 0 | 1 | 0 | 1 | 0 | **10** |
| 0 | 1 | 0 | 0 | 1 | **9** |
| 0 | 1 | 0 | 0 | 0 | **8** |
| 0 | 0 | 1 | 1 | 1 | **7** |
| 0 | 0 | 1 | 1 | 0 | **6** |
| 0 | 0 | 1 | 0 | 1 | **5** |
| 0 | 0 | 1 | 0 | 0 | **4** |
| 0 | 0 | 0 | 1 | 1 | **3** |
| 0 | 0 | 0 | 1 | 0 | **2** |

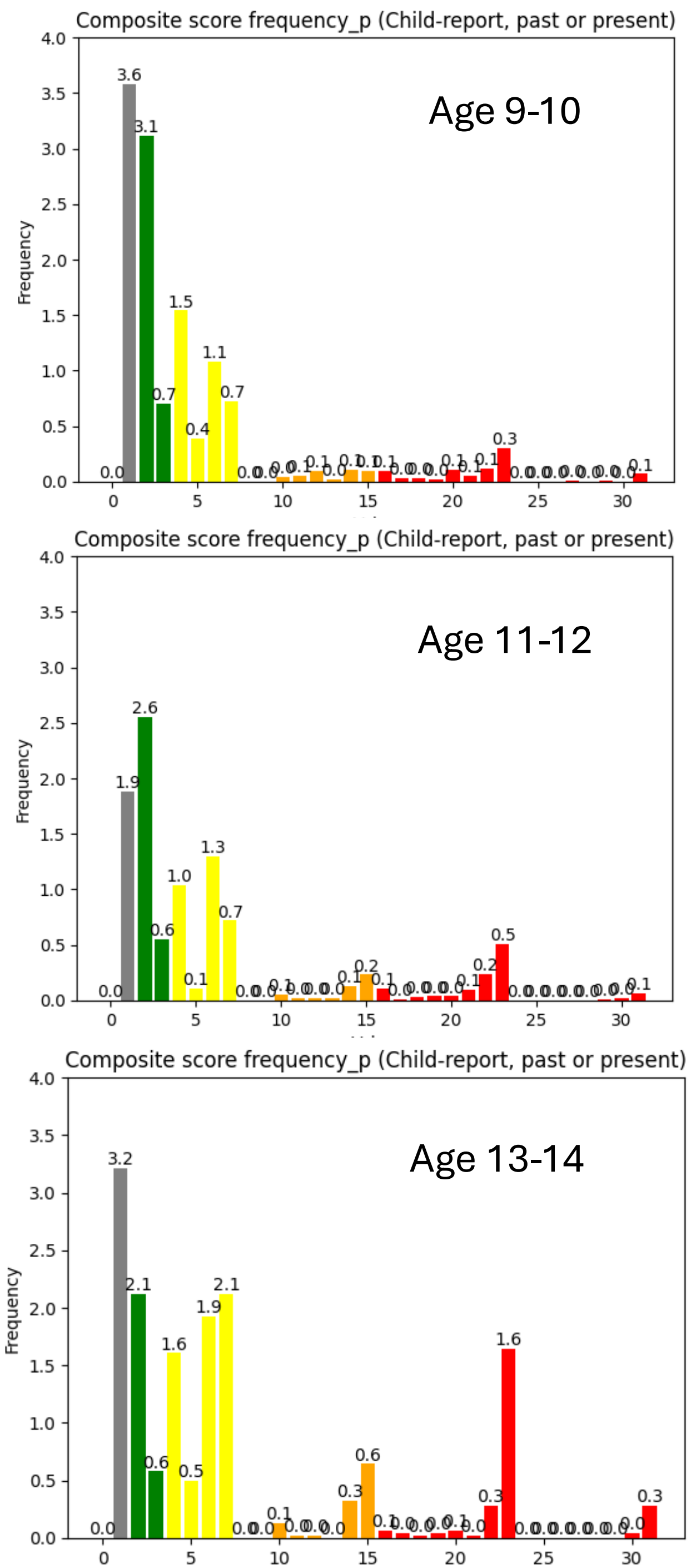

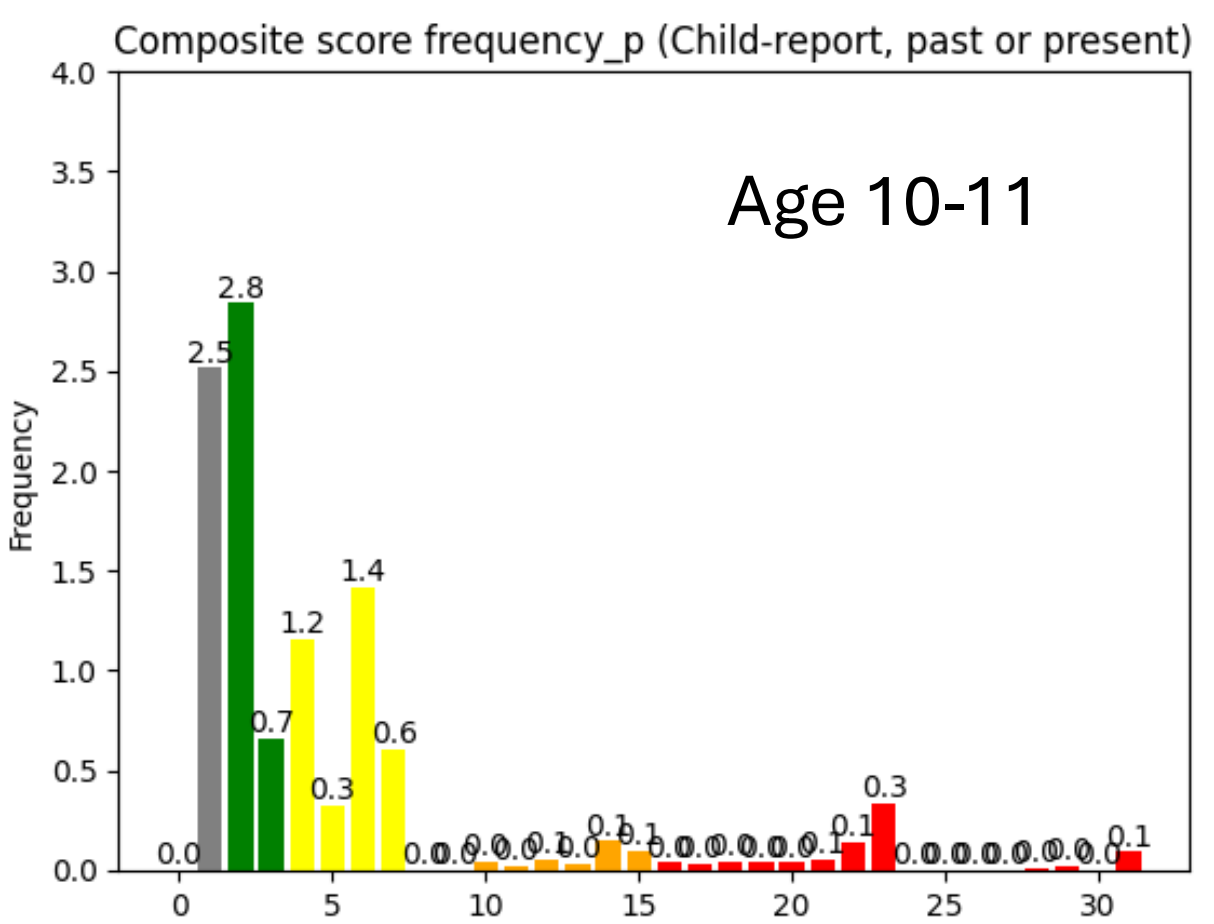

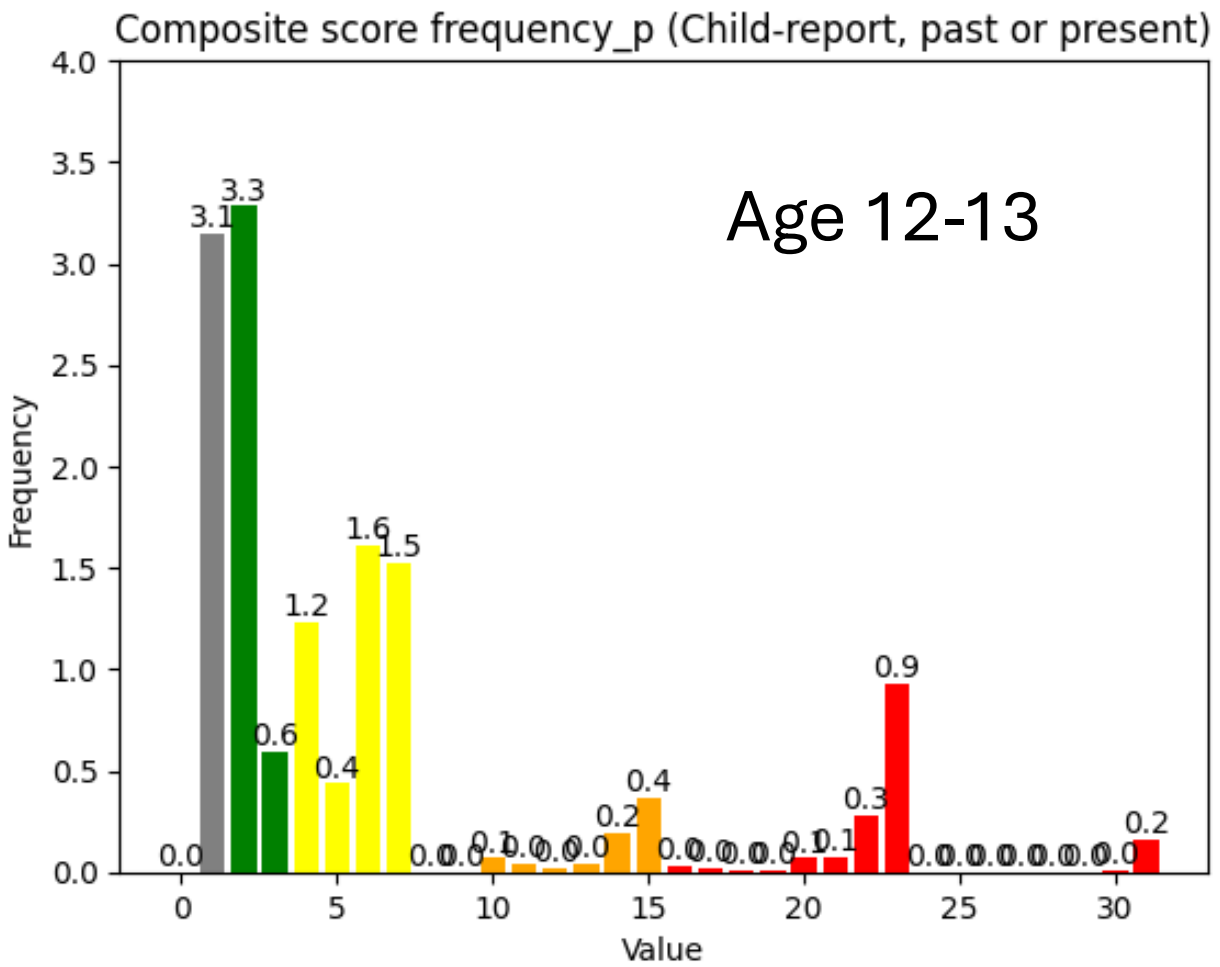

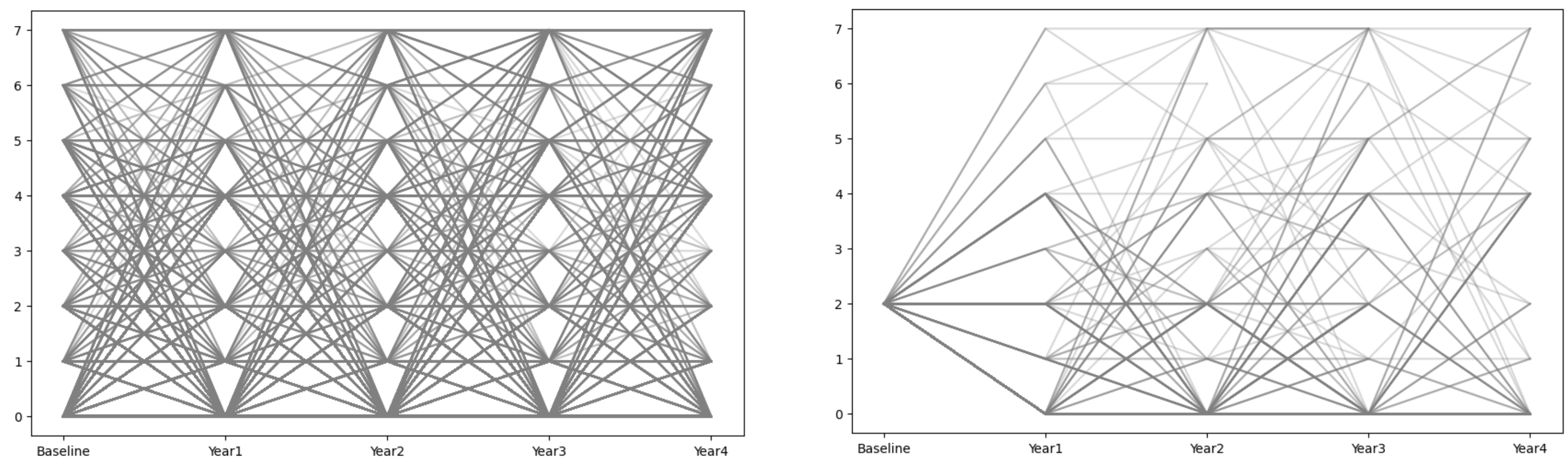

**Extended Data Fig. 2.** Left: Trajectories of all children; Right: Trajectories of children who reported passive ideation (State 2) at baseline (age 9-10). The weight of each line segment indicates the frequency. This illustrates the high variability and fluctuation in the data.